\documentclass[a4paper,11pt]{article}

\usepackage{fullpage}
\usepackage{floatpag}
\usepackage[draft]{fixme}
\usepackage{tikz}
\usepackage{subfigure}
\usepackage{multicol}
\usepackage{bbm}
\usepackage{graphicx}
\usepackage{amsmath}
\usepackage{amsfonts}
\usepackage{amssymb}
\usepackage{amsthm}
\usepackage{verbatim}
\usepackage{mathrsfs}
\usepackage{color,xcolor}
\usepackage{changes}
\usepackage{setspace}
\usetikzlibrary{patterns}

\definecolor{myurlcolor}{rgb}{0,0,0.4}
\definecolor{mycitecolor}{rgb}{0,0.5,0}
\definecolor{myrefcolor}{rgb}{0.5,0,0}
\usepackage[pagebackref]{hyperref}
\hypersetup{colorlinks, linkcolor=myrefcolor,
citecolor=mycitecolor, urlcolor=myurlcolor}

\newcommand{\one}{\leavevmode\hbox{\small1\normalsize\kern-.33em1}}
\def\tr{\mathrm{tr}}

\font\Bbb =msbm10 
 \def\C{{\hbox {\Bbb C}}}

\def\be{\begin{equation}}
\def\ee{\end{equation}}
\def\ben{\begin{eqnarray}}
\def\een{\end{eqnarray}}
\def\eea{\end{array}}
\def\bea{\begin{array}}
\newcommand{\ot}[0]{\otimes}

\newcommand{\bei}{\begin{itemize}}
\newcommand{\eei}{\end{itemize}}
\newcommand{\ket}[1]{|#1\rangle}

\newcommand{\braket}[2]{\langle{#1}|{#2}\rangle}

\renewcommand{\emph}[1]{\textbf{#1}}

\newcommand{\figref}[1]{Figure~\ref{#1}}

\newcommand{\secref}[1]{Section~\ref{#1}}
\newcommand{\appref}[1]{Appendix~\ref{#1}}

\theoremstyle{plain}

\theoremstyle{definition}

\theoremstyle{remark}

\begin{document}

\title{Exploring the Local Orthogonality Principle}
\author{A.~B. Sainz$^{1}$, T. Fritz$^{1,2}$, R. Augusiak$^{1}$, J. Bohr Brask$^{1}$,\\ R.
Chaves$^{1,3}$, A. Leverrier$^{4}$, and A. Ac\'in$^{1,5}$\\[0.5em]
{\it\small $^1$ICFO--Institut de Ciencies Fotoniques, E--08860 Castelldefels, Barcelona, Spain}\\
{\it\small $^2$Perimeter Institute for Theoretical Physics, Waterloo, Ontario, Canada}\\
{\it\small $^3$Institute for Physics, University of Freiburg, D-79104 Freiburg, Germany}\\
{\it\small $^4$Inria, EPI SECRET, B.P. 105, 78153 Le Chesnay Cedex, France}\\
{\it\small $^5$ICREA--Institucio Catalana de Recerca i Estudis
Avan\c{c}ats, E--08010 Barcelona, Spain}}

\date{\today}

\maketitle

\begin{abstract}

Nonlocality is arguably one of the most fundamental and counterintuitive aspects of quantum theory. Nonlocal correlations could, however, be even more nonlocal than quantum theory allows, while still complying with basic physical principles such as no-signaling. So why is quantum mechanics not as nonlocal as it could be? Are there other physical or information-theoretic principles which prohibit this? So far, the proposed answers to this question have been only partially successful, partly because they are lacking genuinely multipartite formulations. In~\cite{FSA}, we have introduced the principle of Local Orthogonality (LO), an intrinsically multipartite principle which is satisfied by quantum mechanics but is violated by non-physical correlations.

Here we further explore the LO principle, presenting additional results and explaining some of its subtleties. In particular, we show that the set of no-signaling boxes satisfying LO is closed under wirings, present a classification of all LO inequalities in certain scenarios, show that all extremal tripartite boxes with two binary measurements per party violate LO, and explain the connection between LO inequalities and unextendible product bases.

\end{abstract}

\setcounter{tocdepth}{1}
\setstretch{0}
\singlespacing

\section{Introduction}

Bell's theorem~\cite{Bell} is arguably one of the most fundamental lessons we have learned about Nature in the past decades. It is a no-go theorem which states that certain natural and seemingly obvious assumptions are incompatible with quantum mechanics. These assumptions are that far-apart observers cannot influence each other instantaneously, that they can choose their respective measurements independently, and that physical quantities have well-established values previous to any measurement. These assumptions impose restrictions (Bell inequalities) on the correlations that the distant parties may obtain. The quantum-mechanical violations of Bell inequalities imply that quantum mechanics violates at least one of these natural assumptions underlying Bell's theorem. This problem is often referred to as \textit{quantum nonlocality}.

Being part of the endeavour to understand the counterintuitive
features of quantum theory as our current most accurate
description of nature, the study of quantum nonlocality has become
an active and fruitful field of research. Several new discoveries
have uncovered both prerequisites and consequences of nonlocality,
revealing an intrinsic interdisciplinarity (see e.g.~our
connections to graph theory) and striking applications in a wide
range of topics like quantum key
distribution~\cite{Acin2007,Pironio2009,VV12b} or certified
randomness generation~\cite{Pironio2010,CK11,VV12a}.

However, despite these successes, understanding the structure of
quantum nonlocality is still a fundamental open problem. No
concise characterization of \textit{which} nonlocal correlations
can arise in quantum mechanics has been found so far, and it is
unclear whether such a characterization even exists. The first
observation in this direction was made by
Tsirelson~\cite{tsirelson} and Popescu-Rohrlich~\cite{pr} who
noticed that the nonlocality of quantum theory is not maximal
within theories satisfying some well-founded physical principles
such as the impossibility of instantaneous communication
(\textit{no-signaling}). They provided paradigmatic examples of
correlations between two parties compatible with the no-signaling
principle, but without any quantum realization. This correlation
has now become known as the \textit{Popescu-Rohrlich box}, or
\textit{PR-box} for short. Given this insufficiency of the
no-signaling principle, the big question is whether there exists
some other fundamental principle explaining the structure of
quantum correlations.

More recently, several principles with an information-processing flavor have been proposed to this end. These include non-trivial communication complexity~\cite{vd,Brassard2006}, Information Causality~\cite{Pawlowski2009a}, and Macroscopic Locality~\cite{nw}. Unfortunately, while being more restrictive than no-signaling, none of these principles can be sufficient to recover the set of quantum correlations exactly. The reason for this lies in the fact that intrinsically multipartite principles are essential for characterizing the set of quantum correlations: there exist supra-quantum correlations for three parties that cannot be witnessed by any bipartite principle~\cite{gallego}, even when the parties are split in two groups and the principle is applied to each such bipartition. Most or all of the existing principles for quantum correlations are formulated in a bipartite setting and it is unclear whether they have more powerful multipartite generalizations.

In order to fill this gap, we have recently introduced Local Orthogonality (LO)~\cite{FSA}, an intrinsically multipartite principle for correlations. This principle is based on a definition of orthogonality of events in Bell scenarios. Here, an \textit{event} consists of a specification of measurement choices $x_1\ldots x_n$ and outcomes $a_1\ldots a_n$. We defined a pair of events to be \textit{orthogonal} whenever these two events involve different outcomes of the same local measurement by at least one of the parties. Imposing that the sum of the probabilities of pairwise orthogonal events is $\leq 1$ implies a restriction on the possible correlations. These are Bell inequalities, and we call them \textit{LO inequalities}. One of our basic observations is that all of them are satisfied by quantum correlations. Violations of LO inequalities hence witness supra-quantum correlations.

The aim of this paper is to work out in detail the implications of
the LO principle introduced in~\cite{FSA}. In
Section~\ref{se:non-locality}, we recall the relevant notions like
Bell scenarios and no-signaling boxes. Section~\ref{lo:def}
introduces the LO principle at the single-copy level. After
showing that it is always satisfied by quantum correlations, we
show how it can be violated by certain no-signaling boxes in
multipartite scenarios. The connections to graph theory that we
explain then are crucial tools for our further investigations. In
Section~\ref{se:LOinfty}, we note that the LO principle is
violated by two copies of a PR-box, which motivates the
introduction of the LO$^\infty$ principle, stating that any
physically realistic box should satisfy LO for any number of
copies. We show that all nonlocal extremal boxes in the $(3,2,2)$
scenario violate the LO$^\infty$ principle. We speculate that
LO$^\infty$ may recover Tsirelson's bound, and explain why proving
this is very difficult. In Section~\ref{se:wirings}, we study how
LO$^\infty$ constrains no-signaling boxes constructed via wirings.
We prove that if wired copies of a given box violate LO$^\infty$, then
so does the original box. Hence, in order to find violations of LO
or LO$^\infty$, it is enough to consider independent copies of the
box. In Section~\ref{se:UPB}, we present a connection between LO
inequalities and unextendible product bases (UPBs), including a
construction that turns an LO inequality into a (weak) UPB. In
Appendix~\ref{sec.equiv}, we explain how LO inequalities can be
classified and show the results of our corresponding computations. 
We provide files with our inequalities for download. 
Finally, in Appendix~\ref{boxpacking}, we show how the question of
whether LO implies Tsirelson's bound can be reformulated in terms
of a box packing problem.

\section{Bell scenarios, nonlocal correlations and no-signaling boxes}
\label{se:non-locality}

This section is an introduction to Bell scenarios and nonlocal correlations which can be skipped by readers familiar with the subject.

It is a well-accepted fact that the predictions of quantum theory are incompatible with those of classical physics \cite{DI}. This was first proven by Bell~\cite{Bell} via a gedankenexperiment in 1964, and since then many actual experiments have corroborated this~\cite{Freedman1972,Aspect1982,Rowe2001,Matsukevich2008}. This result, known as \textit{Bell's theorem}, establishes a very counterintuitive aspect of quantum theory: either the stated assumptions on the causal structure (no-signaling and the possibility to choose measurements independently of the experiment) are wrong, or measurements do not have predetermined outcomes prior to being conducted. This phenomenon is called ``quantum nonlocality''.

In more detail, the setup of Bell's gedankenexperiment is as
follows: two separated parties---commonly called Alice and
Bob---each have access to a physical system. Typically, these two
systems are particles, e.g.~photons, which originate from the same
source and hence may be correlated with each other. In each run of
the experiment, each of Alice and Bob is allowed to freely choose
one among a fixed number $m$ of measurements, each having a fixed
number $d$ of possible outcomes, to perform on their system. The
measurement choices are usually denoted by $x$ for Alice and $y$
for Bob, while the corresponding outcomes are written as $a$ and
$b$. The measurements are arranged so that they define
space-like separated events. If the parties take note of the
outcomes in each run of the experiment and gather statistics, they
will eventually obtain a conditional probability distribution
$P(ab|xy)$. In order to have a more concise terminology, we often
use the term \textit{correlation} to refer to such a conditional
distribution $P(ab|xy)$.

The same gedankenexperiment can be conducted with any number $n$ of ``parties'' in place of Alice and Bob. In this ``multipartite'' case, we denote the measurement choices by $x_1\ldots x_n$, the measurement outcomes by $a_1\ldots a_n$, and the resulting correlation by $P(a_1\ldots a_n|x_1\ldots x_n)$. We take the $m$ possible measurement choices to be given by $x_i\in\{0,\ldots,m-1\}$ and the $d$ possible outcomes by $a_i\in\{0,\ldots,d-1\}$ An individual conditional probability $P(a_1\ldots a_n|x_1\ldots x_n)$ in a correlation represents the probability for the parties to get outcomes $a_1\ldots a_n$ upon performing measurements $x_1\ldots x_n$. The concrete \textit{Bell scenario} in which such a correlation lives is specified by the number of parties $n$, the number of measurements per party $m$, and the number of outcomes per measurement $d$. We simply write $(n,m,d)$ for such a specification; the smallest non-trivial scenario is $(2,2,2)$~\cite{chsh}.

\subsection{Classical correlations}

An explanation of a correlation in terms of classical physics is given by a \textit{local hidden variable} (LHV) model. An LHV model is specified by the following data:
\begin{itemize}
\item a ``hidden variable'' $\lambda$, with probability distribution $q(\lambda)$, which describes the joint state of all systems together,
\item deterministic functions $f_i(x_i,\lambda)$ which, for each party $i$, determines the outcome $a_i$ as a function of the measurement $x_i$ and the joint state $\lambda$.
\end{itemize}
The classical correlation described by such an LHV model is then given by~\cite{Bell,Fine1982}

\be
\label{classical}
P(a_1 \ldots a_n|x_1 \ldots x_n) = \sum_\lambda q(\lambda)\, \delta_{a_1,f_1(x_1,\lambda)}\cdots\delta_{a_n,f_n(x_n,\lambda)}.
\ee

We call a correlation \textit{classical} if it arises in this way. The term \textit{local} is often used as a synonym for classical. It was shown by Fine~\cite{Fine1982} that replacing the functions $f_i(x_i,\lambda)$ with probabilistic assignments $P_i(a_i|x_i,\lambda)$ does not give a bigger set of correlations.

The main point raised by Bell is that there exist correlations which arise from measurements on entangled quantum systems which are not classical. Such correlations are also called {\it nonlocal}~\cite{DI}. But first, what does ``arise from measurements on entangled quantum systems'' mean for a correlation? This is what we explain next.

\subsection{Quantum correlations}

Quantum correlations are those obtained by the parties
when they have access to quantum systems and measurements. The
system shared among the $n$ parties is then represented by an
$n$-partite quantum state, which is described as a density matrix
$\rho$ on a joint Hilbert space $\mathcal{H}=\otimes_{i=1}^n
\mathcal{H}_i$. A measurement $x_i$ by party $i$ on their system
is described by a collection of orthogonal projections
$\Pi_{0}^{x_i},\ldots,\Pi_{d-1}^{x_i}$ on $\mathcal{H}_i$ which
satisfy $\sum_{a_i} \Pi_{a_i}^{x_i} =
\mathbbm{1}_{\mathcal{H}_i}$. The correlation $P(a_1\ldots
a_n|x_1\ldots x_n)$ arises via Born's rule as \be \label{quantum}
P(a_1 \ldots a_n|x_1 \ldots x_n) = \tr_{\mathcal{H}}\!\left(\rho
\,  \Pi_{a_1}^{x_1} \otimes \cdots \otimes \Pi_{a_n}^{x_n}
\right). \ee A correlation is called \textit{quantum} if it arises
in this way. For many entangled states $\rho$, one can find
measurements such that the resulting correlation is not
classical---this is the content of Bell's theorem. On the other
hand, it is not hard to show that every classical correlation is
also quantum.

Equivalently, instead of requiring the $\Pi_{a_i}^{x_i}$
to be projections, one could allow the use of general
measurements, defined by positive-operator-valued measures (POVMs),
in the definition of quantum correlation. Since we do not fix the
dimensionality of the $\mathcal{H}_i$, every quantum correlation
with POVMs can also be realized with projections by adjoining an
ancillary system to each $\mathcal{H}_i$. This is analogous to the
classical case from above, in which the models were restricted to
deterministic functions $f_i$ without loss of generality.

\subsection{No-signaling boxes}
\label{se:box-world}

In this section, we study general correlations which satisfy the {\it no-signaling principle}, that is, correlations which are not in conflict with the impossibility of instantaneous or superluminal influence. This comprises both classical and quantum correlations.

In some situations, it is convenient to avoid any assumptions on the systems the parties have or the specific way the measurements are implemented. This framework is usually called {\it box world}, and is used for the study of {\it device-independent} tasks, such as device-independent quantum key distribution~\cite{Acin2007,Pironio2009,VV12b} and device-independent random number generators~\cite{Pironio2010,VV12a}. Here, each party is thought of as having access to a device (``box'') which produces an outcome when one of several available buttons is pressed. In order for these devices to be well-defined objects, there should exist a correlation $P(a_1\ldots a_n|x_1\ldots x_n)$ for each box, which represents its observed behavior. The no-signaling principle constrains these correlations to be those which satisfy the \textit{no-signaling equations}

\begin{equation}\label{NS-cond}
\sum_{a_{i}} P(a_1\ldots a_i  \ldots  a_n |x_1  \ldots x_i \ldots x_n) = \sum_{a_i} P(a_1\ldots a_i  \ldots a_n |x_1 \ldots x'_i \ldots  x_n )  .
\end{equation}

This ensures that all marginals of $P$ are well-defined, i.e.~do not depend on the measurement choices of the parties that are marginalized over. In any given scenario $(n,m,d)$, those $P$ which satisfy the no-signaling equations are the inhabitants of box world.

In the following, we use the term ``no-signaling box'', or simply ``box'', for a no-signaling correlation. A box which is not quantum is also called \textit{supra-quantum}.

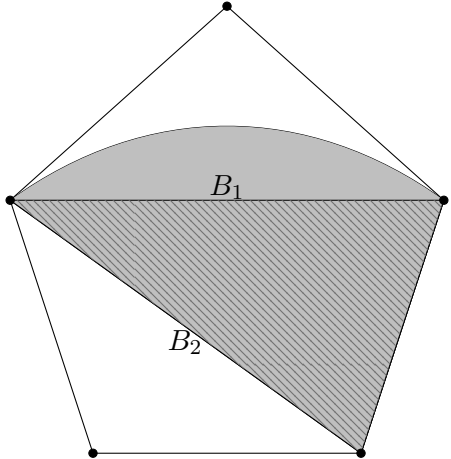
\begin{figure}
\begin{center}
\begin{tikzpicture}
    \draw (162:3) to [out=36,in=144] (18:3) -- (306:3) -- cycle;
    \draw (18:3) -- (90:3.5) -- (162:3) -- (234:3) -- (306:3) -- cycle;
    \path[fill=lightgray] (18:3) to [out=144,in=36] (162:3) -- (306:3) -- cycle;
    \draw (18:3) -- (162:3) -- (306:3) -- cycle;
    \path[pattern=north west lines, pattern color=gray!90!black] (18:3) -- (162:3) -- (306:3) -- cycle;
    \node[draw,shape=circle,fill,scale=.3] (a) at (18:3) {} ;
    \node[draw,shape=circle,fill,scale=.3] (b) at (90:3.5) {} ;
    \node[draw,shape=circle,fill,scale=.3] (c) at (162:3) {} ;
    \node[draw,shape=circle,fill,scale=.3] (d) at (234:3) {} ;
    \node[draw,shape=circle,fill,scale=.3] (e) at (306:3) {} ;
    \node at (90:1.1) {$B_1$};
    \node at (240:1.1) {$B_2$};
\end{tikzpicture}
\end{center}
\caption{Schematic representation of the sets of no-signaling
correlations (outer pentagon), quantum correlations (gray area)
and classical correlations (striped area). The lines $B_1$ and
$B_2$ separating the set of classical correlations from the
nonlocal ones are examples of tight Bell inequalities. While $B_1$
is violated by some quantum correlations, $B_2$ is only violated
by supra-quantum boxes. In fact, $B_2$ is an example of a
Bell inequality that is tight, as it corresponds to a facet of the
set of classical correlations, non-trivial, because it is violated
by no-signaling correlations, but with no quantum violation.}
\label{fig:poly}
\end{figure}

See \figref{fig:poly} for a schematic representation of the three sets of correlations. Both the sets of classical and no-signaling correlations are polytopes. The set of quantum correlations is still convex, but has a more subtle geometry: generally, its boundary has both flat parts and curved parts.

\section{Local Orthogonality, level $1$}
\label{lo:def}

Here we introduce and discuss the principle of Local Orthogonality, as it applies to one copy of any no-signaling box. In the upcoming sections, we show that LO is much more powerful when applied to many independent copies of a box and study the ramifications of doing this. But for now, we keep things simple and consider the single-copy case only.

\subsection{Orthogonal events in Bell scenarios}

An \textit{event} in a Bell scenario $(n,m,d)$ is what a correlation assigns a probability to: it is the specification $(a_1\ldots a_n|x_1\ldots x_n)$ consisting of a particular joint outcome $a_1\ldots a_n$ and a particular choice of measurements $x_1\ldots x_n$. In~\cite{FSA}, we called a pair of events
\be
\label{lopair}
e=(a_1 \ldots a_n|x_1 \ldots x_n),\qquad e'=(a'_1\ldots a'_n|x'_1\ldots x'_n)
\ee
\textit{locally orthogonal}, or simply \textit{orthogonal}, if they involve different outcomes of the same measurement by (at least) one party, that is, if $a_i \neq a_i'$ and $x_i=x_i'$ for some $i$. Intuitively, this means that the two events are \textit{exclusive}: upon fixing a particular value of $\lambda$ in an LHV model~\eqref{classical}, at most one of these two events can occur, since $\delta_{a_i,f(x_i,\lambda)}=1$ means that $a_i=f(x_i,\lambda)$, and therefore $a'_i\neq f(x_i,\lambda)$. This implies that $P(e)+P(e')\leq 1$ for any classical correlation $P$.

More generally, we call a collection of events $\{e_j\}$ orthogonal if these events are pairwise orthogonal. The same reasoning applies to show that, for a fixed value of $\lambda$, at most one of these events can occur in any given LHV model. Therefore, any classical correlation necessarily satisfies the inequality
\be
\label{loineq}
\sum_j P(e_j) \leq 1.
\ee
Writing this out in terms of $e_j = (a^j_1\ldots a^j_n|x^j_1\ldots x^j_n)$ gives the same equation in more explicit form,
\be
\sum_j P(a^j_1\ldots a^j_n|x^j_1\ldots x^j_n) \leq 1.
\ee
Since we have shown such an inequality to be valid for all classical correlations, we are actually dealing with a Bell inequality. The Bell inequalities of this form are our \textit{LO inequalities}. We therefore arrive at~\cite{FSA},
\begin{quote}
\textbf{Local Orthogonality principle:}
Any physically realistic correlation must satisfy all LO inequalities. In other words, for any collection of orthogonal events, the sum of their conditional probabilities must not be larger than one.
\end{quote}
To summarize: the LO principle (i) introduces a notion of orthogonality between two events, (ii) makes a set of events jointly orthogonal whenever they are pairwise orthogonal, and (iii) requires the inequality~(\ref{loineq}) to be satisfied for any jointly orthogonal set of events.

This differs from other principles such as Information Causality in the sense that it is not based on a statement for two parties and then extended to multipartite scenarios by applying it to any bipartition. On the contrary, the LO principle is intrinsically multipartite.

It has been stated~\cite{cabello} that the LO principle ``follows {\textup{[\,\dots]}} from Boole's axiom of probability stating that the sum of the probabilities of events that are jointly exclusive cannot exceed $1$''. This statement is incorrect and constitutes an invalid application of Boole's axiom, since the probabilities in~\eqref{loineq} are \textit{conditional} on the given measurements, and this condition is typically different for each term in the inequality. In order to apply Boole's axiom, one needs to make the probabilities in a box $P(a_1\ldots a_n|x_1\ldots x_n)$ unconditional by choosing probabilities $P(x_1\ldots x_n)$ for the measurements and constructing the associated joint distribution of measurements \textit{and} outcomes as
\[
P(a_1\ldots a_n,x_1\ldots x_n) = P(a_1\ldots a_n|x_1\ldots x_n)\cdot P(x_1\ldots x_n).
\]
Now even if the original box violates~\eqref{loineq}, this equation results in a well-defined probability distribution and, in particular, satisfies Boole's axiom. See also~\cite{Henson} for a related discussion.

So if it is not Boole's axiom, then what does the LO principle mean, intuitively? This is not so easy to answer, and we refer to the ``distributed guessing'' interpretation in~\cite{FSA} for one proposed solution to this problem.

\subsection{LO and quantum correlations}\label{se:qcorr}

Do quantum correlations necessarily satisfy the LO principle? As we explain now, this is indeed the case~\cite{FSA}, and this is what has raised our interest in it.

For a quantum correlation as in~\eqref{quantum}, every event $e=(a_1\ldots a_n|x_1\ldots x_n)$ has an associated operator given by
\[
\Pi_e := \Pi^{x_1}_{a_1}\otimes\ldots\otimes\Pi^{x_n}_{a_n},
\]
and it is easy to show that this is a projection. We then have the simple formula
\be
\label{Borne}
P(e)=\tr_{\mathcal{H}}(\rho\Pi_e).
\ee
Our crucial observation is that if two events $e$ and $e'$ are orthogonal, then the respective projections are also orthogonal in the sense that $\Pi_e \, \Pi_{e'} = \Pi_{e'} \, \Pi_{e} = 0$. Now for a set $\{e_j\}$ of orthogonal events, the pairwise orthogonality of the projections implies that
\[
\sum_j \Pi_{e_j} \leq \mathbbm{1}.
\]
Using~\eqref{Borne}, we therefore arrive at
\[
\sum_j P(e_j) = \tr_\mathcal{H} \left( \rho \sum_j \Pi_{e_j}\right) \leq \tr_\mathcal{H} (\rho) = 1,
\]
which means that the given quantum correlation satisfies the given LO inequality. In other words, quantum correlations satisfy the LO principle.

Despite the simplicity of this argument, we see that the LO
principle---suitably extended to the many-copy level as explained
in \secref{se:LOinfty}---bounds the set of quantum correlations
surprisingly tightly. It is a nontrivial property of quantum
mechanics that pairwise orthogonality of a set of projections
implies ``joint exclusivity'' in the sense of having a sum upper
bounded by $1$. This property has also been considered in
the context of contextuality.  For instance, several works have
studied how the violation of the corresponding orthogonality
conditions can lead to supra-quantum correlations~\cite{specker,
spekkens}, or how it can be used to provide upper bounds to the
quantum violation of non-contextuality inequalities~\cite{csw}.

%

\subsection{LO and no-signaling correlations}\label{se:lo&ns}

Now that we know that quantum correlations satisfy the LO
principle, it is time to ask whether the same is true for no-signaling
correlations. Or is the principle
violated by some no-signaling, and therefore supra-quantum, correlations?

First of all, for a pair of orthogonal
events as in~\eqref{lopair}, the LO inequality $P(e)+P(e')\leq 1$
holds for any no-signaling box $P$. This is because the two events
$e$ and $e'$ can be seen as different outcomes of a correlated
measurement in which party $i$ first measures $x_i$ and announces
the outcome to the other parties, and then the other parties apply
measurements depending on this outcome: any party $j\neq i$
applies measurement $x_{j}$ if $i$'s outcome was $a_i$, and $x'_j$
otherwise. Then, since $e$ and $e'$ are outcomes of this
correlated measurement, exclusiveness of outcomes
implies that $P(e) + P(e') \leq 1$. Note that the no-signaling
principle is essential for this correlated measurement to be
meaningful, as it makes it possible to talk about the probabilities for
$i$'s outcome independently of the chosen measurements of the
other parties. See the discussion of the Foulis-Randall product
in~\cite{FLS} for further considerations along these lines. This
result can be extended to any set of pairwise orthogonal events in
all bipartite scenarios, as proven in~\cite{FSA,csw}. The LO
principle (at the currently discussed first level) is then
equivalent to the no-signaling principle.

All the previous results suggest that the set of LO
correlations coincides with the no-signaling set. If this was the
case, then the LO principle would be completely useless and not
constrain the physically realizable correlations any better than
the no-signaling principle already does. But fortunately, it was
shown in~\cite{FSA} that the equivalence between the LO and
no-signaling principles breaks down when considering larger sets
of orthogonal events in multipartite Bell scenarios. It is
precisely this non-equivalence in the multipartite scenario that
makes the LO principle non-trivial. As becomes more clear
in what follows, this non-equivalence is also the reason behind
the activation effects in the application of LO, which explains why
LO is non-trivial even in the bipartite case.

To prove the non-equivalence between LO and the
no-signaling principle, it is enough to consider the simplest
multipartite scenario $(3,2,2)$. An interesting LO inequality in
this scenario is the {\it Guess-Your-Neighbor's-Input} (GYNI)
inequality~\cite{gyni}, \be \label{gyni}
P(000|000)+P(110|011)+P(011|101)+P(101|110) \leq 1. \ee It is easy
to see by simple inspection that GYNI is an LO inequality, since
all events in this sum are pairwise orthogonal. As shown
in~\cite{gyni}, the maximal value of the left-hand side of this
inequality over no-signaling boxes is $4/3$, which proves the
existence of no-signaling boxes violating LO. An example of an LO
violation which is easier to check by hand is given in
\secref{se:LOinfty}.

More generally, all the examples of tight Bell inequalities without quantum violations given in~\cite{gyni,UPBBell1,UPBBell2,AAABbook} are actually examples of LO inequalities.

\subsection{LO and graph theory}
\label{se:lo&gt}

Given that there are no interesting LO inequalities in a bipartite
scenario, and in addition in multipartite scenarios it is not so easy to
write any of    them down at all, the question arises: how does
one go about finding some or even all LO inequalities in a given
scenario? Here we give a complete answer to this question, using
the language and tools from graph theory. This
construction is inspired by the graph approach to contextuality
introduced in~\cite{csw}.

\begin{figure}
\centering
\includegraphics[width=0.5\textwidth]{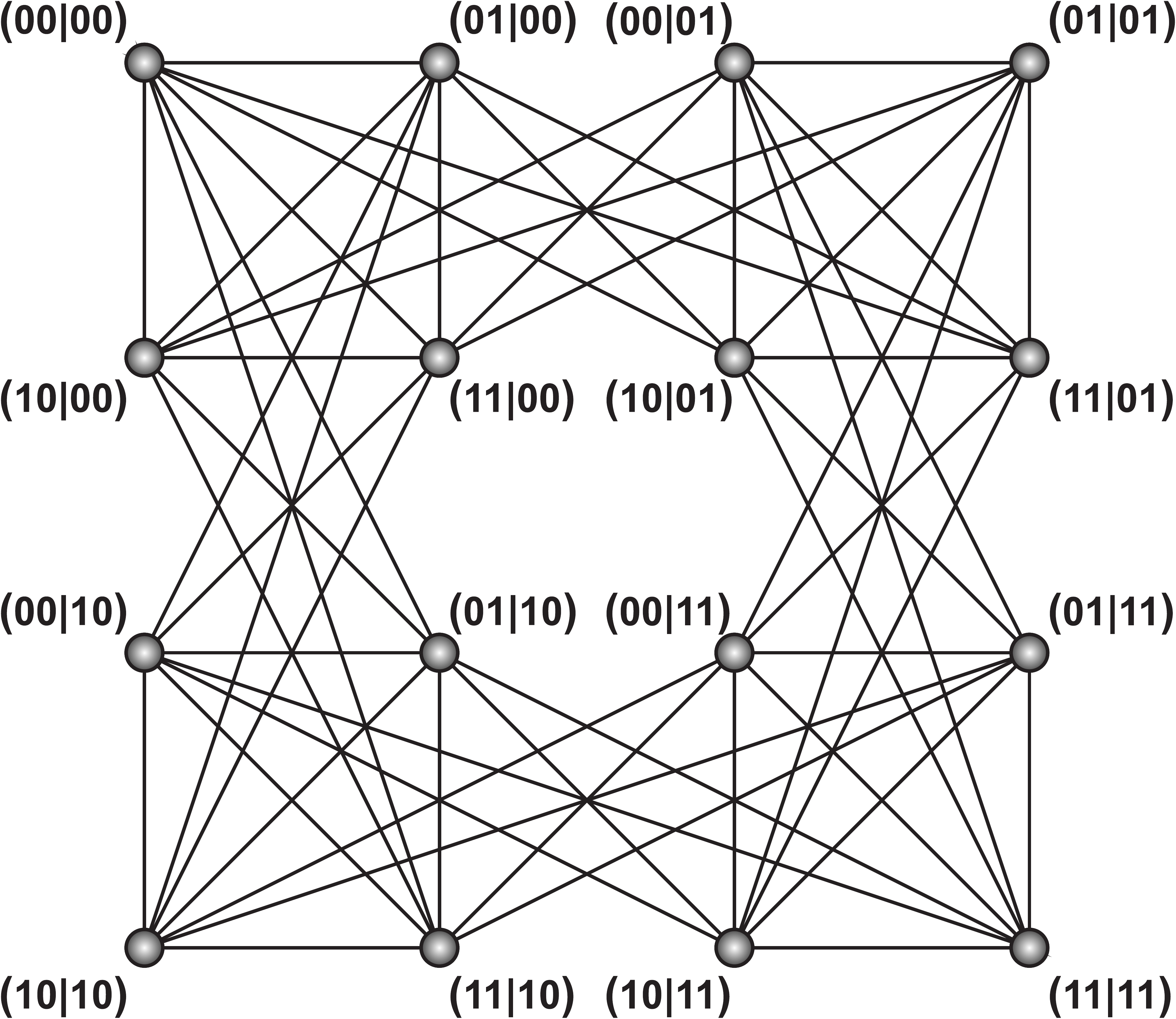}
\caption{Orthogonality graph of the $(2,2,2)$ scenario. As
mentioned in the text, each possible event corresponds to a node,
while the edges connect locally orthogonal events.} \label{OG2}
\end{figure}

The \textit{orthogonality graph} $O_{n,m,d}$  associated with a Bell scenario $(n,m,d)$ is defined as follows. We take its vertices to correspond to the events of the scenario; hence there are $(md)^n$ vertices. Two vertices are connected by an edge if and only if the corresponding events are orthogonal. For instance, \figref{OG2} shows the orthogonality graph of the $(2,2,2)$ scenario.

In graph theory, a \textit{clique} in a graph $G$ is a subset of vertices $C \subseteq V(G)$ such that the subgraph induced by $C$ is complete, i.e.~such that all pairs of vertices in $C$ are connected by an edge in $G$. A clique is {\it maximal} if it cannot be extended to another clique by including a new vertex. By our earlier definition of orthogonality, a set of events in a Bell scenario is orthogonal if and only if the corresponding vertices in the orthogonality graph form a clique. In this way, a clique in the orthogonality graph gives rise to an LO inequality, and vice versa. Moreover, if one clique $C$ is contained in another clique $C'$, then the LO inequality associated with $C'$ is better than the one associated to $C$, since it contains more terms on the left-hand side. For this reason, maximal cliques correspond to \textit{optimal} LO inequalities to which no further term can be added on the left-hand side, and it is sufficient to work only with these when applying the LO principle.

In conclusion, the problem ``find all the optimal LO inequalities'' is equivalent to ``find all maximal cliques in the orthogonality graph''. Although the problem of finding all maximal cliques of a graph is known to be NP-hard\footnote{Note, however, that in principle, while the problem of finding the maximal cliques is NP-hard for general graphs, this may no longer be the case for the subset of graphs that arise as the orthogonality graphs of Bell scenarios.}~\cite{clique_nphard}, there exist software packages~\cite{mace,cliquer} which can find all maximal cliques in sufficiently small graphs. We have used these packages to find all LO inequalities in various Bell scenarios (see \appref{sec.equiv}).

How does one now find out whether a given box $P$ satisfies the LO principle? This can be answered by a variation of the very same methods. If one thinks of the probability $P(a_1\ldots a_n|x_1\ldots x_n)$ of an event $e=(a_1\ldots a_n|x_1\ldots x_n)$ as a number, or \textit{weight}, assigned to the corresponding vertex in $O_{n,m,d}$, then the box violates some LO inequality if and only if there is a clique of total weight $>1$. Fortunately, the software packages~\cite{mace,cliquer} can also handle such weighted graphs and answer this kind of question by computation. It is enough to include only those events in the graph which have a non-zero probability to occur for the given box, since vertices of weight zero do not contribute at all to the total weight of a clique. Making use of this fact has simplified some of our computations significantly.

It is particularly simple to use these weighted graphs for boxes for which all non-zero probabilities are equal. After having discarded events with probability zero, all vertices in the graph---corresponding to the possible events---have the same weight attached to them; let us call this weight $c$. Then there exists a clique of total weight $>1$ if and only if there exists a clique of size $>c^{-1}$. Now asking whether there exists a clique of size larger than a given number has nothing to do with the weights, and we can again consider the corresponding graph as unweighted. The results of our computations with these methods are presented in the next section.

\begin{figure}
\centering \subfigure[{}]{\includegraphics[scale=0.3]{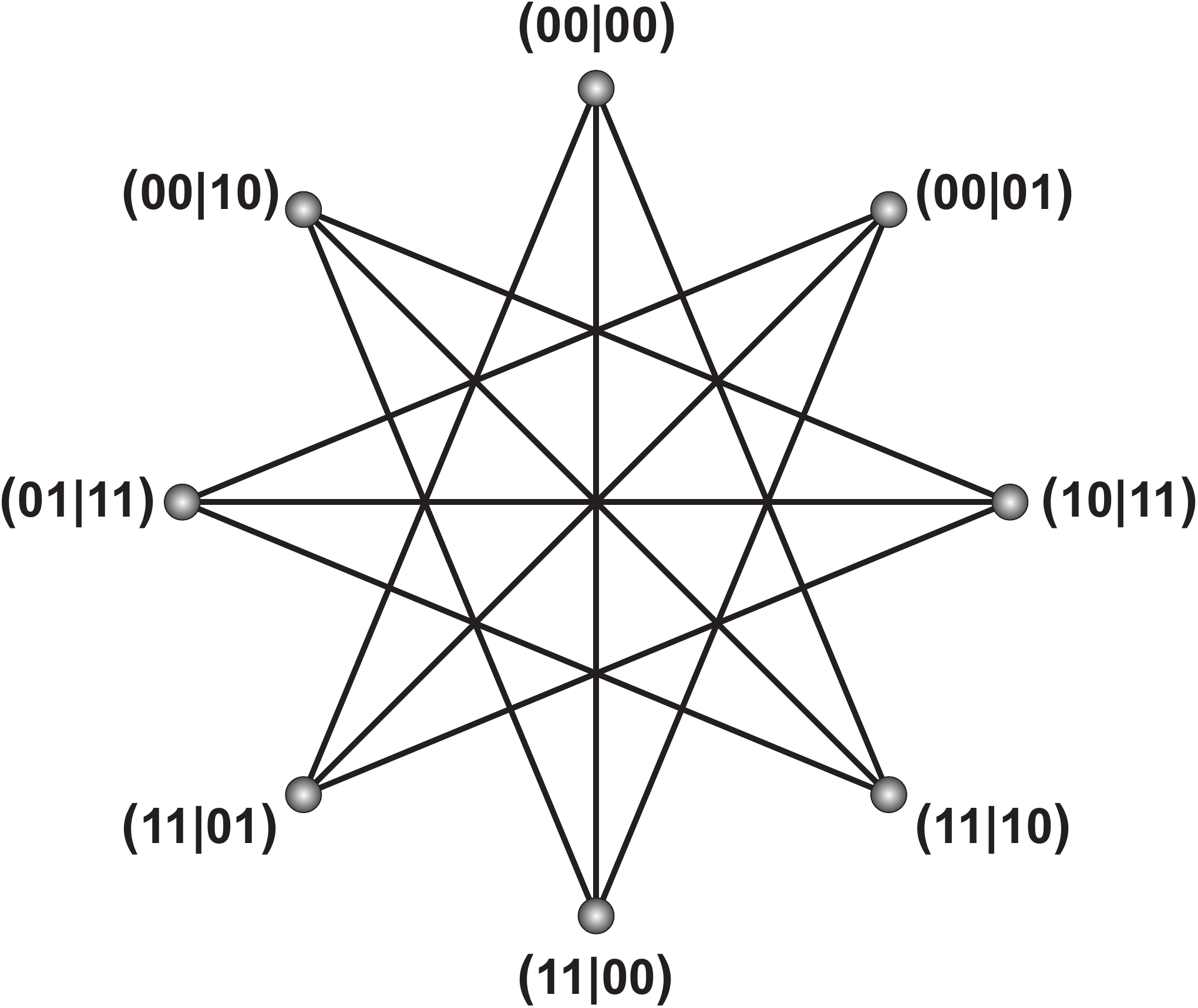}
\label{grafoPR}}\hspace{2cm}
\subfigure[{}]{\includegraphics[scale=0.25]{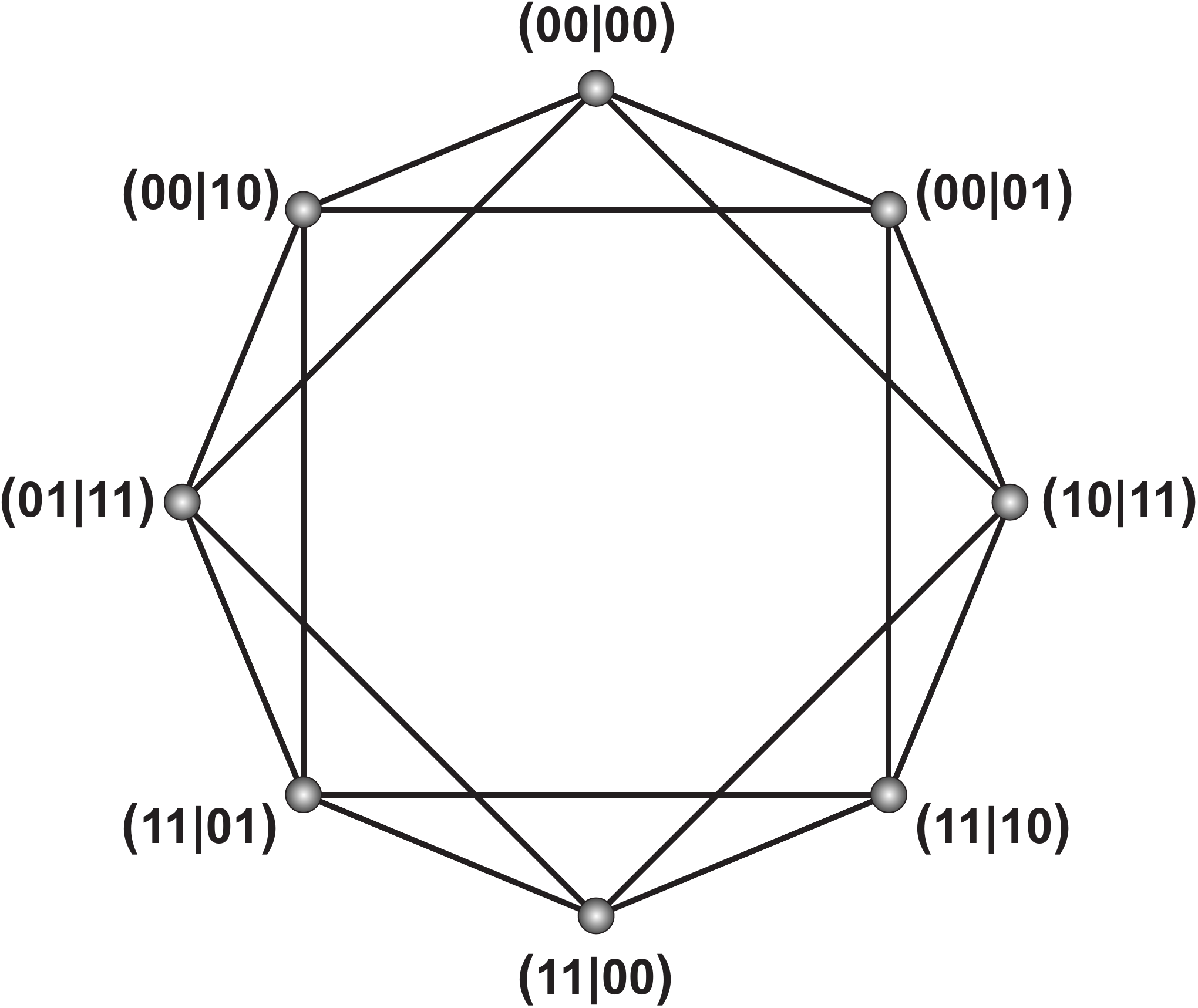}
\label{circulant}} \caption{\subref{grafoPR} Orthogonality Graph
of possible events for a PR-box. It coincides with Figure~2
in~\cite{SBBC}, where the authors study the CHSH inequality.
\subref{circulant} Non-orthogonality graph of the PR-box, i.e.~the
complement of~\subref{grafoPR}. In graph theory terms, this is the
circulant graph $Ci_8(1,2)$ and also the $4$-antiprism
graph~\cite{GL}.} \label{PRgraphs}
\end{figure}

A final remark about graphs: unfortunately, in order to connect with the mathematical literature which uses opposite conventions, we sometimes need to work in terms of the \textit{non-orthogonality graph} $\mathrm{NO}_{n,m,d}$~\cite{FLS}, which is simply defined as the graph complementary to the orthogonality graph: its vertices are again the $(md)^n$ events of the Bell scenario, and two such events share an edge if they are \textit{not} orthogonal. The same reasoning as in the previous two paragraphs shows that optimal LO inequalities then correspond to maximal independent sets, where an \textit{independent set} in a graph is a subset of vertices which do not share any edges at all.

\section{Local Orthogonality, level $\infty$}
\label{se:LOinfty}

If a certain principle like LO is supposed to single out those boxes that are physically realistic, then the set of boxes satisfying the principle needs to be closed under certain operations that are physically realizable. This comprises operations like \textit{wirings}, which we consider in \secref{se:wirings} in complete generality, and in particular situations as simple as considering several independent copies of box rather than just one copy. The obvious question then is whether a number $k$ of copies of an $n$-partite box $P$, forming the $nk$-partite box $P^{\otimes k}$, necessarily satisfies LO if the original box $P$ does. We show now that this is not necessarily the case: LO can be \textit{activated} by applying it to multiple copies of a box jointly. This leads us to define an infinite hierarchy of LO principles.

\subsection{A hierarchy of LO principles}\label{PR:vio}

In principle, one might think that because of the equivalence between no-signaling and the LO principle in the bipartite case, LO might be useless for the detection of supra-quantum bipartite boxes. For example, the PR-box in the $(2,2,2)$ scenario is given by the conditional probability distribution
\begin{equation}\label{PR-pos}
{PR}(ab|xy) = \begin{cases} \frac{1}{2} & \text{if }a \oplus b =xy, \\ 0 & \text{otherwise}. \end{cases}
\end{equation}
This box is known to be maximally nonlocal in the sense that it reaches the algebraic maximum for the Clauser-Horne-Shimony-Holt (CHSH) inequality~\cite{chsh}, thereby violating the Tsirelson bound of quantum correlations~\cite{tsirelson2}.
\begin{figure}
\centering
\includegraphics[width=0.5\linewidth]{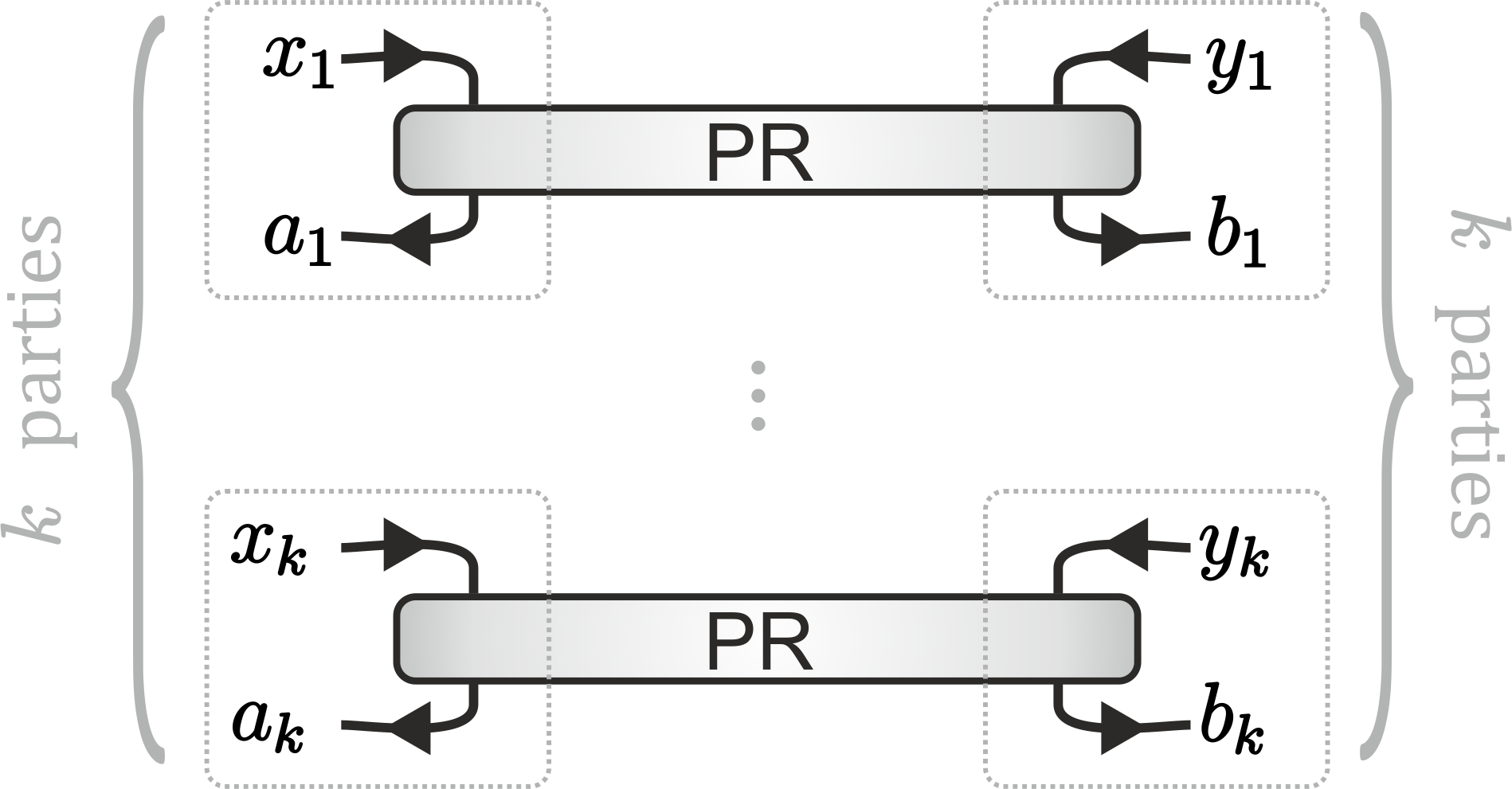}
\caption{$k$ copies of a PR-box shared among $2k$ parties. Each
party has access to one part of one box.} \label{f-manyPR-1}
\end{figure}
For $k$ independent copies of a PR-box, distributed among $2k$ parties as shown in \figref{f-manyPR-1}, the conditional probability distribution is
\begin{equation}
\label{def-NPR}
PR^{\otimes k}(a_1 b_1 \cdots a_k b_k | x_1y_1 \cdots x_k y_k ) = \prod_{j=1}^k PR(a_j b_j | x_j y_j) .
\end{equation}
Now already for $k=2$, we find LO inequalities violated by these two copies of the PR-box. One such inequality is
\be
\label{PRviol}
P(0000|0000)+P(1110|0011)+P(0011|0110)+P(1101|1011)+P(0111|1101)\leq 1,
\ee
with a left-hand side evaluating to $5/4$ for $P=PR^{\otimes 2}$.

This shows two things: first, the LO principle is not stable under
taking copies of a box; second, this phenomenon can be exploited
to witness the supra-quantumness of some bipartite boxes, despite
the equivalence between the LO principle and no-signaling
at the one-copy level.

We now move on to postulating LO also on the many-copy level. For every number $k$, we therefore obtain the following:
\begin{quote}
\textbf{LO$^k$ principle:}
For any physically realistic correlation $P$, the original LO principle must hold for $P^{\otimes k}$: for any set of orthogonal events in the $kn$-partite scenario, the sum of their conditional probabilities must not be larger than one.
\end{quote}
After all, if a box $P$ is physically realizable, then so is $P^{\otimes k}$ for any $k$, and therefore also $P^{\otimes k}$ should satisfy the LO principle. Since there is no need to limit this reasoning to any particular value of $k$, we are led to the following:
\begin{quote}
\textbf{LO$^\infty$ principle:}
For any physically realistic correlation $P$, the original LO principle must hold for any $P^{\otimes k}$: for any $k$ and any set of orthogonal events in the $kn$-partite scenario, the sum of their conditional probabilities must not be larger than one.
\end{quote}

We write $\mathcal{LO}^k$ for the set of correlations that satisfy the LO$^k$ principle, and $\mathcal{LO}^\infty$ for those that satisfy the LO$^\infty$ principle. These sets satisfy a chain of inclusions:
$$
\mathcal{LO}^\infty \subseteq \ldots \subseteq \mathcal{LO}^k \subseteq
\mathcal{LO}^{k-1} \subseteq \ldots \subseteq \mathcal{LO}^1.
$$
We have $\mathcal{LO}^\infty=\cap_k \mathcal{LO}^k$ by definition. This set is the most interesting one in our hierarchy, but also the most difficult one to describe.

Exploiting the whole hierarchy of principles, LO$^\infty$ allows us to detect the non-quantumness of bipartite correlations: as we have seen, already the LO$^2$ principle is stronger than the no-signaling principle in bipartite scenarios. On the other hand, since the set of quantum correlations is closed under taking tensor products $P\mapsto P^{\otimes k}$, quantum correlations necessarily satisfy LO$^\infty$ as well.

Using the graph-theoretical methods explained in \secref{se:lo&gt}, we can systematically search for violations of LO$^k$ by the PR-box, or by any other given no-signaling box, for any fixed $k$. More precisely, finding a violation of LO$^k$ by the PR-box requires finding an LO inequality in the $(2k,2,2)$ scenario which is violated by $PR^{\otimes k}$. This means that we need to look for maximal cliques in the orthogonality graph $O_{2k,2,2}$, which has $16^k$ vertices. However, since only $8^k$ of the events in this scenario have non-zero probability in $PR^{\otimes k}$, it is sufficient to search for maximal cliques in the subgraph induced by these $8^k$ possible events. The PR-box violates LO$^k$ if and only if this graph has a clique of size greater than $2^k$. As demonstrated by~\eqref{PRviol}, there exist cliques of size $5>2^2$ in $O_{4,2,2}$. Our computations showed that this is the only violation of LO$^2$ by the PR-box up to symmetry.

\subsection{LO$^\infty$ and supra-quantum tripartite boxes}\label{se:tribox}

In the previous subsection, we proved that the LO$^\infty$ principle witnesses the PR-box---the only extremal bipartite correlation in the $(2,2,2)$ scenario---as unphysical. In the next section, we show also that all extremal no-signaling boxes in the $(2,2,d)$ and $(2,m,2)$ scenarios are likewise ruled out by LO. So what about boxes in multipartite scenarios?

We have applied the methods of \secref{se:lo&gt} to all the extremal boxes in the $(3,2,2)$ scenario, which had been computed in~\cite{tri-box}. These fall into $46$ equivalence classes under symmetries, with one class corresponding to deterministic local boxes, while the other $45$ ones are nonlocal. How do the LO and LO$^\infty$ principles perform on these? Table~\ref{t-box} shows the results: \textit{all} of these boxes violate LO$^1$ or LO$^2$. We have found this by first searching for a violation of LO$^1$ for any given box; for those that did not display any violation of LO$^1$, we used the same methods to look for violations of LO$^2$, and successfully found such violations in all cases.

Since the GYNI inequality~\eqref{gyni} is the only LO inequality in the $(3,2,2)$ scenario up to symmetry (\secref{sec.equiv}), those boxes that violate LO$^1$ do actually violate GYNI.

\begin{table}
\begin{multicols}{3}
\begin{tabular}{c|c|c|c}
box & copies & terms & value \\
\hline
2 & 2 & 5 & $5/4$ \\
3 & 2 & 17 & $17/16$ \\
4 & 2 & 17 & $17/16$ \\
5 & 1 & 4 & $5/4$ \\
6 & 2 & 17 & $17/16$ \\
7 & 2 & 17 & $17/16$ \\
8 & 2 & 9 & $9/8$ \\
9 & 1 & 4 & $9/8$ \\
10 & 1 & 4 & $9/8$ \\
11 & 1 & 4 & $5/4$ \\
12 & 2 & 17 & $9/8$ \\
13 & 1 & 4 & $7/6$ \\
14 & 1 & 4 & $7/6$ \\
15 & 1 & 4 & $9/8$ \\
16 & 1 & 4 & $7/6$ \\
\end{tabular}

\begin{tabular}{c|c|c|c}
box & copies & terms & value \\
\hline
17 & 1 & 4 & $7/6$ \\
18 & 1 & 4 & $9/8$ \\
19 & 2 & 17 & $17/16$ \\
20 & 1 & 4 & $6/5$ \\
21 & 2 & 17 & $17/16$ \\
22 & 2 & 19 & $37/36$ \\
23 & 1 & 4 & $7/6$ \\
24 & 1 & 4 & $7/6$ \\
25 & 1 & 4 & $4/3$ \\
26 & 2 & 22 & $37/36$ \\
27 & 2 & 17 & $17/16$ \\
28 & 1 & 4 & $7/6$ \\
29 & 1 & 4 & $4/3$ \\
30 & 2 & 14 & $26/25$ \\
31 & 2 & 13 & $26/25$ \\
\end{tabular}

\begin{tabular}{c|c|c|c}
box & copies & terms & value \\
\hline
32 & 1 & 4 & $6/5$ \\
33 & 1 & 4 & $6/5$ \\
34 & 2 & 13 & $37/36$ \\
35 & 1 & 4 & $7/6$ \\
36 & 2 & 13 & $33/32$ \\
37 & 1 & 4 & $9/8$ \\
38 & 1 & 4 & $7/6$ \\
39 & 2 & 22 & $37/36$ \\
40 & 2 & 17 & $17/16$ \\
41 & 2 & 17 & $17/16$ \\
42 & 2 & 17 & $17/16$ \\
43 & 2 & 14 & $50/49$ \\
44 & 2 & 17 & $17/16$ \\
45 & 2 & 17 & $17/16$ \\
46 & 2 & 17 & $17/16$
\end{tabular}
\end{multicols}
\caption{Table of LO violations for the extremal tripartite boxes
in the numbering scheme of~\cite{tri-box}. The columns show the
number of copies needed for finding the violation, the number of
terms in the violated inequality and the value given by the box
for that inequality. Since it is always larger than 1, all these
boxes violate some LO inequality.} \label{t-box}
\end{table}

The intrinsically multipartite character of the LO principle has
allowed us to witness the supra-quantumness of box number $4$
in~\cite{tri-box}, which cannot be achieved with any bipartite
principle~\cite{3box-IC}. One example of an LO inequality in the
$(6,2,2)$ scenario which is violated by two copies of this box is
given by
\begin{align*}
& \quad\: P(000000|000000) + P(000101|000100) + P(000010|011101) + P(000111|011001)
\\[5pt]
& + P(001110|110001) + P(010110|110101) + P(011011|010010) + P(011111|010001)
\\[5pt]
& + P(100000|011100) + P(101111|010111) + P(101100|100111) + P(110100|000101)
\\[5pt]
& + P(110110|000001) + P(110011|010111) + P(111010|000100) + P(111100|000111)
\\[5pt]
& \hspace{10.46cm} + P(111000|110100) \leq 1.
\end{align*}

\subsection{LO$^\infty$, noisy boxes and Tsirelson's bound}
\label{S_capacities}

In this subsection, we consider noisy versions of the PR-box and
other boxes and ask how robust the violations of LO$^\infty$ are
with respect to noise. At a certain critical purity level, the
LO$^\infty$ violation of a given box disappears, while at another
(possibly identical) purity level, the box becomes quantum. We
relate the first critical purity level to the \textit{Shannon
capacity} of graphs in a manner similar to that in~\cite{FLS}, although
the details of this connection are different. We had already shown
in~\cite{natcomm} that LO$^\infty$ detects the non-quantumness of
many bipartite boxes, including some for which all other known
principles like Information Causality have so far failed to do so.

Since LO$^2$ detects the PR-box as unphysical, it is interesting to see to what extent this also applies to noisy PR-boxes. These are boxes indexed by a purity parameter $q\in[0,1]$,
\[
{PR}_{q} := q \,PR + (1-q) P_{\mathbb{I}},
\]
where
\[
P_\mathbb{I}(a b|x y):=1/4 \qquad \forall a,b,x,y
\]
is the maximally noisy box. The value $q=1$ corresponds to the PR-box, $q=1/\sqrt{2}$ corresponds to the statistics obtained when measuring a singlet state with the appropriate measurements (leading to a saturation of the Tsirelson bound), and $q=1/2$ is a local box giving the maximal classical value for the CHSH inequality. Any box in the $(2,2,2)$ scenario can be turned into one of this form by local operations without changing its value of the CHSH inequality~\cite{MAG}. Since these local operations are particular instances of wirings, the results of Section~\ref{se:wirings} show that it is enough to show that any $PR_q$ with $q>1/\sqrt{2}$ violates LO$^\infty$ in order to show that LO$^\infty$ recovers Tsirelson's bound.

We focus first on the two-copy case, $k=2$, meaning that we consider the box
\[
{PR}_q^{\otimes 2}(a_1 b_1 a_2 b_2|x_1y_1x_2y_2) = \prod_{i=1}^2 \left( q \, PR(a_ib_i|x_iy_i)
+ (1-q) P_\mathbbm{I}(a_ib_i|x_iy_i) \right).
\]
For $q<1$, all events in the underlying $(4,2,2)$ scenario have non-zero probability, and hence we cannot discard any vertices in the orthogonality graph. The clique corresponding to the inequality~\eqref{PRviol} is not maximal, but it can be completed to a maximal clique in one or several ways. By doing so, we obtain inequalities with additional terms which vanish for $PR^{\otimes 2}$, but do not vanish for ${PR}_q^{\otimes 2}$. A direct computation shows that for all $q>(\sqrt{10}-1)/3\approx 0.721$, the noisy box violates the $10$-term LO inequality
\begin{align*}
& \quad\: P(1111 | 0000) + P(1100 | 1010) + P(0100 | 1100) + P(0011 | 0001) + P(0010 |0111)
\\[5pt]
& + P(1011 | 0000) + P(0101 | 1100) + P(1101 | 1100) + P(1010 |0110) + P(1001 | 0100) \leq 1.
\end{align*}
This purity level $q\approx 0.721$ is remarkably close to
Tsirelson's bound $q=1/\sqrt{2}\approx 0.707$, and it is tempting
to conjecture that the optimal $q$ converges to Tsirelson's bound
as the number of parties increases. Our computations have shown
that this is the optimal value that can be derived with $k=2$, and
we have not found any improvement for $k=3$.

It is a very difficult problem to determine whether the critical
purity level as $k\to\infty$ coincides exactly with Tsirelson's
bound. To see \textit{how} difficult it is, we now explain how it
is related to the problem of computing the Shannon capacity of a
certain graph. Similar considerations can be found
in~\cite{cabello}. For many graphs, computing the Shannon capacity
is a notoriously hard combinatorial problem~\cite{AL}. While we
explain the following for the concrete example of the PR-box, the
very same considerations apply to any no-signaling box for which
all non-zero probabilities are equal.

If we want to determine whether a noisy box $PR_q^{\otimes k}$ violates some LO inequality, we need to check whether there exists an independent set in the non-orthogonality graph $\mathrm{NO}_{2k,2,2}$ to which $PR_q^{\otimes k}$ assigns a total weight of more than $1$. This is a very difficult problem in general, and therefore we consider an approximation to it. The idea is that those events which contribute the most to the value of an LO inequality are those which already have non-zero probability in the noiseless $PR^{\otimes k}$. This is true for an event
\[
e=(a_1b_1\ldots a_kb_k|x_1y_1\ldots x_ky_k)
\]
in the $(2k,2,2)$ scenario if and only if each $e_i:=(a_ib_i|x_iy_i)$ has non-zero probability in the original $PR$, and all such joint events have the same probability $2^{-k}$. In this way, we can identify the vertices of the non-orthogonality graph $\mathrm{NO}_{PR^{\otimes k}}$ with the $k$-tuples of vertices in the single-copy graph $\mathrm{NO}_{PR}$ displayed in \figref{PRgraphs}.

Similarly, two events $e$ and $e'$ of this form are orthogonal if and only if they are orthogonal at some $i$. So two vertices in $\mathrm{NO}_{PR^{\otimes k}}$, thought of as $k$-tuples of vertices in $\mathrm{NO}_{PR}$, share an edge if and only if for every component index $i$, their $i$-components either coincide or share an edge in $\mathrm{NO}_{PR}$. In graph-theoretic terms, these two properties constitute the definition of the $k$-fold \textit{strong product} $\mathrm{NO}_{PR}^{\boxtimes k}$, so that we obtain
\[
\mathrm{NO}_{{PR}^{\otimes k}} = \mathrm{NO}_{PR}^{\boxtimes k}.
\]
In fact, analogous reasoning shows that $O_{n,m,d}$ is the $n$-fold co-normal product of $O_{1,m,d}$, where $(1,m,d)$ is the ``Bell scenario'' with only one party. See~\cite{FLS} for more detail and applications of this observation.

A maximal independent set in $\mathrm{NO}_{{PR}^{\otimes k}}$ corresponds to an LO inequality in scenario $(2k,2,2)$. However, this inequality may not be optimal in the sense that it may be tightened by adding further events which have probability zero for the noiseless box ${PR}^{\otimes k}$, but non-zero probability for a noisy box ${PR}^{\otimes k}_q$. This is the approximation we make: looking only at this subclass of LO inequalities does not give optimal bounds, but it is necessary for making the connection to the Shannon capacity of $NO_{PR}$.

We write $q_k$ for the maximum value of $q$ for which ${PR}_q$ satisfies the LO$^k$ principle. By what we just said, this critical purity level $q_k$ is upper bounded by the maximum value of $q$ for which ${PR}^{\otimes k}_q$ satisfies all LO inequalities corresponding to maximal independent sets in $\mathrm{NO}_{PR}^{\boxtimes k}$. We denote this alternative critical purity level by $q^*_k$.

All events in $\mathrm{NO}_{PR}^{\boxtimes k}$ carry the same probability, whose value is given by
\[
\left(\frac{1+q}{4}\right)^k.
\]
Therefore ${PR}_q^{\otimes k}$ satisfies all LO inequalities coming from $\mathrm{NO}_{PR}^{\boxtimes k}$ if and only if
\[
\left(\frac{1+q}{4}\right)^k\cdot \alpha_k \leq 1,
\]
where $\alpha_k$ denotes the size of a maximal independent set in $\mathrm{NO}_{PR}^{\boxtimes k}$. This gives the critical purity level
\be
\label{qbound}
q_k \leq q^*_k = \frac{4}{\sqrt[k]{\alpha_k}}-1.
\ee
For example for $k=2$, we found earlier that $\alpha_2 = 5$, and hence $q^*_2 = 4/\sqrt{5}-1 \approx 0.789$, which is consistent with our previous observation that $q_2\approx 0.721$.

The reason that this is interesting is because the term $\sqrt[k]\alpha_k$ is intimately related to a graph invariant of $\mathrm{NO}_{PR}$, namely its \textit{Shannon capacity}. In fact, the very definition of the Shannon capacity is given by
\[
\Theta(\mathrm{NO}_{PR}) := \lim_k \sqrt[k]{\alpha_k}.
\]
Naturally, this definition applies to any other graph in place of $\mathrm{NO}_{PR}$. For more on the Shannon capacity and its relation to nonlocality and contextuality, including the standard proof showing that the limit over $k$ exists, we refer to~\cite{FLS}.

In terms of the Shannon capacity, we can turn~\eqref{qbound} into an upper bound on the critical purity level at which $PR_q$ just satisfies LO$^\infty$,
\be
\label{c2p}
q_\infty \leq q^*_\infty = \frac{4}{\Theta(\mathrm{NO}_{PR})} - 1.
\ee
Now since $PR_{1/\sqrt{2}}$ is a quantum box and therefore satisfies LO$^\infty$, we know that $q_\infty\geq 1/\sqrt{2}$, which translates by~\eqref{c2p} into $\Theta(\mathrm{NO}_{PR}) \leq 4(2-\sqrt{2})$. This is a bound which coincides with the Lov{\'a}sz number\footnote{To see that this Lov\'asz number is indeed $4(2-\sqrt{2})$, apply Theorem 1 in~\cite{BCCL} with $n=8$.} $\vartheta(\mathrm{NO}_{PR})$, a graph invariant known to upper bound the Shannon capacity.

The LO$^\infty$ principle recovers Tsirelson's bound if and only if $q_\infty=1/\sqrt{2}$. Again by~\eqref{c2p}, this would follow from the hypothetical equation $\Theta(\mathrm{NO}_{PR}) \stackrel{?}{=} \vartheta(\mathrm{NO}_{PR}) = 4(2-\sqrt{2})$. Now one might hope that it should be known whether the Shannon capacity of a graph as simple as $\mathrm{NO}_{PR}$ coincides with its Lov{\'a}sz number. Alas, to the best of our knowledge, this is not the case, and finding this out is a very difficult problem. See also Appendix~\ref{boxpacking} for an illustration of how computing the $\alpha_k$ and therefore also $\Theta(\mathrm{NO}_{PR})$ can be interpreted as a box packing problem.

These observations immediately generalize to other boxes $P$ in arbitrary scenarios $(n,m,d)$. Assume that a given box $P$ has the property that each probability $P(e)$ for every event $e$ either vanishes or has a constant value $c>0$. Then, the method we just described for the PR-box can be applied here as well, and we obtain a non-orthogonality graph $\mathrm{NO}_P$ which is an induced subgraph of $\mathrm{NO}_{n,m,d}$. An analogous relation between the critical purity level $q^*_\infty$ and the Shannon capacity $\Theta(\mathrm{NO}_P)$ follows,
\[
\left(q_k\cdot c + (1-q_k)\cdot\frac{1}{d^n}\right)^k \cdot \alpha_k = 1 \qquad\Longrightarrow\qquad  q_\infty^* = \frac{d^n - \Theta(\mathrm{NO}_P)}{(d^n c -1)\Theta(\mathrm{NO}_P)}
\]
So on the one hand, if the Shannon capacity
$\Theta(\mathrm{NO}_P)$ happens to be known, then an
upper bound on the critical purity level $q_\infty\leq q_\infty^*$ can be derived. On the other hand,
lower bounds on $q^*_\infty$ follow from finding quantum representations of
the box at a certain purity level, and these bounds translate into upper
bounds on $\Theta(\mathrm{NO}_P)$, and hence might be of interest for graph theory. However, to be fair, it seems likely that these bounds are dominated by the Lov{\'a}sz number $\vartheta(\mathrm{NO}_P)$.

These considerations should make clear that computing the boundary of $\mathcal{LO}^\infty$ is difficult, and even approximating it is computationally costly. We have seen that this problem is intimately related to a purely combinatorial problem, namely the computation of the Shannon capacity of certain graphs. This is an interesting connection to graph theory, similar to those already found in~\cite{csw, FLS}. Among other things, we have used these connections in~\cite{FLS} to give a conceptually simple proof of Navascu{\'e}s' observation that LO$^\infty$ is not only satisfied by quantum correlations, but more generally by all those correlations satisfying the level $1+AB$ of the NPA hierarchy of semidefinite programs~\cite{NPA0,NPA}.

\section{Wirings are useless}\label{se:wirings}

Wirings~\cite{wirings} are operations that can be applied to one or more boxes in order to produce a new box. For example, a wiring may consist of party $1$ communicating their outcome $a_1$ to party $2$, who uses this outcome as their choice of measurement and obtains an outcome $a_2$. One can then define a new no-signaling box upon identifying parties $1$ and $2$ with a new joint party with joint measurement choice $x_1$ and joint outcome $a_2$. There are many variations and generalizations of this which we will discuss in the following. Our result is that if $k$ copies of a box violate LO$^1$ when wired together, then the same $k$ copies violate LO$^1$ as independent copies without any wiring, meaning that the original single-copy box violates LO$^k$. This means that in order to find violations of LO$^\infty$, considering wirings is ``useless'' in the sense that no additional violations can be found. Equivalently, we show that if a no-signaling box $P$ satisfies LO$^k$, then any other box which can be obtained from $P$ by applying wirings to copies of $P$ distributed among many parties will satisfy LO$^1$. Since our wirings are more general than those of~\cite{wirings}, this result implies that $\mathcal{LO}^{\infty}$, as a set of boxes, is closed under wirings in the sense of~\cite{wirings}. Possibly counterintuitively, we obtain this result \textit{without} showing convexity of $\mathcal{LO}^\infty$, and actually suspect that this set is not convex in at least some scenarios. This suggests that a set of boxes may be closed under wirings despite not being convex.

\begin{figure}
\begin{center}
\includegraphics[scale=.9]{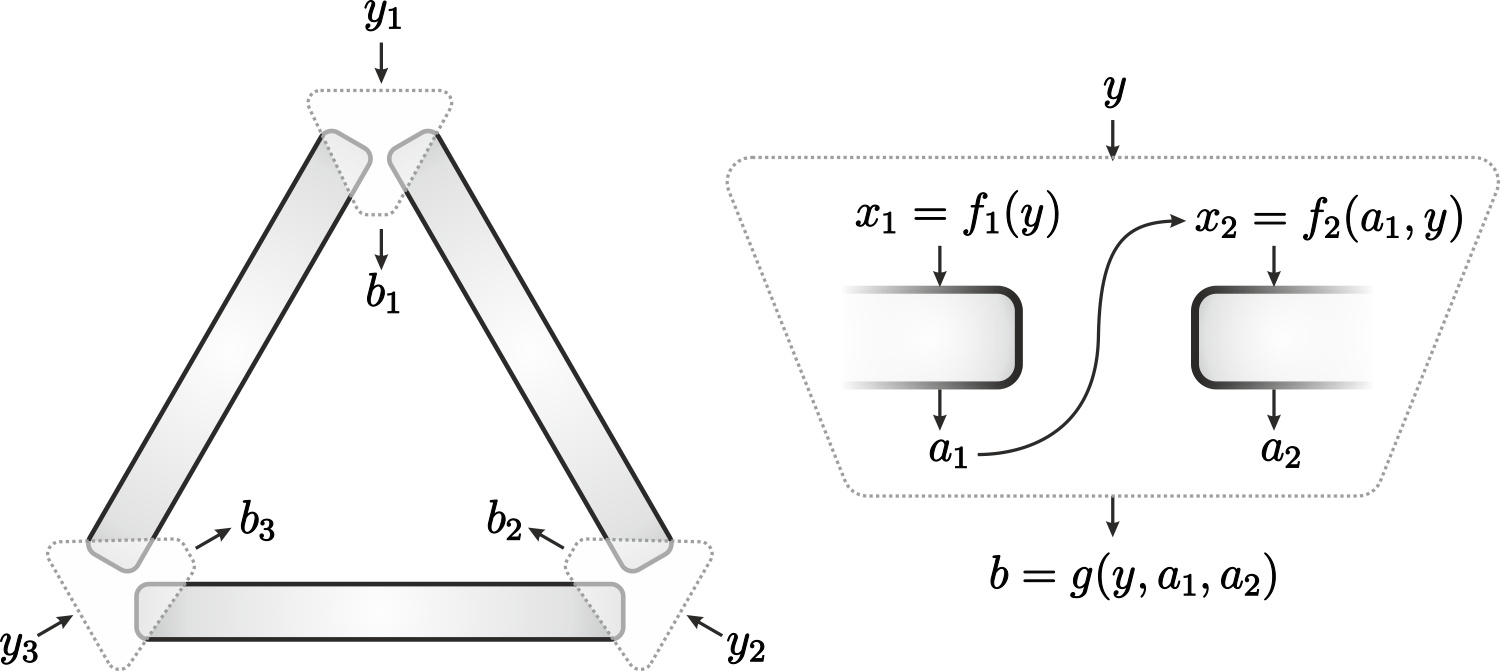}
\end{center}
\caption{Three copies of a bipartite box wired between three
groups of two parties each. The resulting wired box is
tripartite, where each new party has access to one end of two
boxes. To provide an example, consider a grouped party receiving
input $y$. They choose to use first the box on their left-hand
side, to which they input $f_1(y)$ and obtain an outcome $a_1$.
Then, they use the box on the right-hand side by inputting $f_2(y,
a_1)$ and obtaining an outcome $a_2$. Finally, they output
$b=g(y,a_1,a_2)$ as the final outcome of the protocol. Here,
$f_1$, $f_2$ and $g$ can be arbitrary functions.} \label{exwiring}
\end{figure}

We start by considering a restricted, but especially instructive, class of wirings and generalize later. We call these wirings \textit{static} since the ``wires'' are fixed throughout the protocol rather than dynamically positioned, and \textit{deterministic} since no additional randomness other than what comes out of the boxes is involved. A static and deterministic wiring protocol for an $n$-partite box $P$ is obtained as follows: first, we distribute $r$ copies of $P$ among $rn$ parties, where $r$ is arbitrary. This results in the box $P^{\otimes r}$. These $rn$ parties may now assemble into $s$ groups; these groups will become the parties of the ``wired'' box
$$
P_{\mathrm{wired}}(b_1\ldots b_s|y_1\ldots y_s) ,
$$
which they are about to construct. Here, $y_i$ denotes the input which all parties in group $i$ receive \textit{jointly}, while $b_i$ stands for the joint output that they are going to obtain. We may intuitively think of the parties composing a group as physically meeting up at the same location, where each party carries their part of the box to which they have access; see \figref{exwiring} for an example situation. In this way, a particular group formed by $l$ parties has access to $l$ input-output devices.

Second, there is a subprotocol for each group which specifies how the parties within that group communicate and coordinate the use of their boxes. For the sake of concreteness, let us assume that the first group is formed by parties $1,\ldots,l$. For these $l$ parties, we now need to specify an ordering among them, corresponding to the temporal order in which the parties use their devices. For notational convenience, we relabel the parties in the group such that this total ordering is precisely given by the enumeration $1,\ldots,l$.  Party $1$ starts by choosing a measurement choice $f_1(y_1)$ on their device, where $f_1$ is any function which is part of the specification of the protocol, and obtains an outcome $a_1$. Afterwards, party $2$ continues by inputting $f_2(y_1,a_1)$ into their device where $f_2$ is likewise some fixed function, and gets an outcome $a_2$.  In general, party $j+1$ operates after party $j$ and uses the measurement choice $f_{j+1}(y_1,a_1,\ldots,a_j)$ on their device, getting an outcome $a_{j+1}$. After all parties have used their devices in this way, the group announces the total outcome $g(y_1,a_1,\ldots,a_l)$, where $g$ is again a fixed function. In general, such a subprotocol exists for the parties within each group separately.

A trivial example of such a static wiring protocol consists of taking $k=1$, having each party form their own group, and letting each such party apply a deterministic function to their inputs and outputs. This type of wiring corresponds to the special case in which $P_\mathrm{wired}$ is obtained from $P$ via deterministic \textit{local operations}.

We now proceed to show that if $P$ satisfies LO$^\infty$, then so does any $P_{\mathrm{wired}}$ obtained from such a $P$ via static and deterministic wirings. In other words, the set $\mathcal{LO}^\infty$ is \textit{closed under wirings}.

First of all, it is sufficient to show that the wired box satisfies LO$^1$. The reason is that if a box $P_{\mathrm{wired}}$ can be constructed from $P$ via wirings, then so can any of its powers $P_{\mathrm{wired}}^{\otimes k}$, even if the required number of copies of $P$ increases $k$-fold. Our upcoming proof that all boxes constructed from $P$ via wirings satisfy LO$^1$ applies in particular to all the $P_{\mathrm{wired}}^{\otimes k}$. Therefore, $P_{\mathrm{wired}}^{\otimes k}$ satisfies LO$^1$ for all $k$, which means by definition that $P_{\mathrm{wired}}$ satisfies LO$^\infty$.

In order to show that $P_{\mathrm{wired}}$ satisfies LO$^1$, it is enough to consider the case where only the first $l$ parties form a non-trivial group and apply a non-trivial wiring. The reason is that the same argument can be applied to the resulting box, and another non-trivial group can be formed, and this argument can be repeated until all the desired groups have been formed. Assuming this and using the notation from above, the resulting wired box is $s$-partite with $s=rn-l+1$. We enumerate the parties in such a way that the non-trivial group contains the parties $1,\ldots,l$. We make this whole assumption to keep things conceptually simple and not to clutter our notation.

In terms of the above protocol, the conditional probability distribution $P_{\mathrm{wired}}$ of such a wiring of $k$ boxes has the form
\begin{align}
\label{wiredbox}
\begin{split}
& P_{\mathrm{wired}} (b_1\ldots b_s|y_1\ldots y_s) \\[6pt]
&\: = \sum_{\substack{a_1,\ldots a_l\\[1pt] \textrm{s.t.~} g(a_1,\ldots,a_l) = b_1}} P^{\otimes r} (a_1\ldots a_l b_2 \ldots b_s | f_1(y_1) f_2(y_1,a_1) \ldots f_l(y_1,a_1,\ldots,a_l) \,y_2\ldots y_s) .
\end{split}
\end{align}
For fixed $b_1$ and $y_1$, all events occurring in this sum are orthogonal: for any two different terms in the sum represented by indices $a_1,\ldots,a_l$ and $a'_1,\ldots,a'_l$, respectively, let $i$ be the smallest party index for which $a_i\neq a'_i$. Then the two measurement choices of party $i$ are both equal to $f(y_1,a_1,\ldots,a_{i-1})$, while the outcomes are different. This witnesses orthogonality.

Now consider a given set of mutually orthogonal events $(b_1\ldots b_r|y_1\ldots y_r)$ which represents an LO inequality for $P_{\mathrm{wired}}$. We claim that upon substituting~\eqref{wiredbox} into this inequality, we obtain an LO inequality for $P^{\otimes r}$. Checking this means that we need to consider a pair of events that may occur in such an inequality,
\begin{align}
\label{ortevents}
\begin{split}
(a_1\ldots a_l b_2 \ldots b_s | f_1(y_1) f_2(y_1,a_1) \ldots f_l(y_1,a_1,\ldots,a_l) \,y_2\ldots y_s) ,\\[5pt]
(a'_1\ldots a'_l b'_2 \ldots b'_s | f_1(y'_1) f_2(y'_1,a'_1) \ldots f_l(y'_1,a'_1,\ldots,a'_l) \,y'_2\ldots y'_s) .
\end{split}
\end{align}
The case where $b_j=b'_j$ and $y_j=y'_j$ for all $j$, i.e.~where the two events occur in the same sum~(\ref{wiredbox}), was already considered above, where we found them to be orthogonal. Otherwise, there exists some $j$ for which $y_j = y'_j$ and $b_j \neq b'_j$, due to the assumption that the original inequality for $P_{\mathrm{wired}}$ is a LO inequality. If $j\in\{2,\ldots,s\}$, then the two events~(\ref{ortevents}) are clearly orthogonal as well. If $j=1$, then there has to exist some index $i$ for which $a_i\neq a'_i$; upon considering the smallest $i$ with this property, we again find the events~(\ref{ortevents}) to be orthogonal in the same manner as above. Now due to the assumption that $P^{\otimes r}$ satisfies all LO inequalities, we conclude that also $P_{\mathrm{wired}}$ satisfies the given LO inequality.

This shows that the set $\mathcal{LO}^\infty$ is closed under static deterministic wirings. In other words, if a wired box violates LO$^\infty$, then so does the original box from which it was constructed.

We now generalize to ``dynamic'' deterministic wirings in which the temporal ordering of the parties within a group is itself determined during the execution of the protocol. Again we take parties $1,\ldots,l$ to form the only non-trivial group. After receiving their input $y_1$, the party which measures their box first is given by a function $i_1(y_1)$. This party $i_1(y_1)$ performs the measurement $x_{i_1} = f_{1}(y_1)$ and obtains an outcome $a_{i_1}$. This outcome, together with the initial input, determines the second party in the protocol to be the party $i_2(y_1, a_{i_1})$. Similarly, this party then chooses the measurement $x_{i_2} = f_{2}(y_1, a_{i_1})$ and obtains an outcome $a_{i_2}$. The third party in the protocol then is $i_3(y_1, a_{i_1}, a_{i_2})$, and so on. When all parties have finished, the group announces their joint outcome $g(y_1,a_{i_1},\ldots,a_{i_l})$.

In the case of such a dynamic wiring, the explicit form of the sum in~\eqref{wiredbox} is considerably messier to write explicitly and we refrain from doing so. Nevertheless, all events occurring in the corresponding sum for fixed $b_j$ and $y_j$ also satisfy the property of being orthogonal. Indeed, consider two events $e$ and $e'$ in this sum. Both events originated by party $i_1(y_1)$ applying a measurement. Now consider the temporally first step $t$ of the wiring protocol at which the protocol realizations of $e$ and $e'$ differ. Since the protocols are deterministic except for the randomness in the boxes, this difference of the realizations must originate from the previous step $t-1$ by one box having produced different outcomes, $a_{i_{t-1}} \neq a'_{i'_{t-1}}$, although the parties were the same, $i'_{t-1}=i_{t-1}$, and the measurements were the same, $x_{i_{t-1}} = x'_{i'_{t-1}}$. Hence, the two events $e$ and $e'$ are orthogonal. All other statements which we made for static wirings apply directly to dynamic wirings as well, and $\mathcal{LO}^\infty$ is in particular also closed under dynamic deterministic wirings.

So far, all the wirings that we have considered have been \textit{deterministic}: no randomness is allowed in the protocols in the sense that the functions $f_j$, $i_j$ and $g$ are required to be deterministic. So do our results still hold if we do allow randomness in the protocols?

In order to answer this question in the most elegant way, we consider wirings not of copies $P^{\otimes r}$ of the original box $P$, but of any box of the form $P':=P^{\otimes r}\otimes P_{\mathrm{loc}}$, where $P_{\mathrm{loc}}$ is any local box. Again we assume that $P$ satisfies LO$^\infty$, and in particular $P^{\otimes r}$ satisfies LO$^1$. Since $P_{\mathrm{loc}}$ is a convex combination of deterministic boxes all of which satisfy LO$^1$, $P'$ also is a convex combination of boxes which satisfy LO$^1$, and hence $P'$ satisfies LO$^1$ as well.

The point of this consideration is that the local box $P_{\mathrm{loc}}$ may provide shared and/or local randomness for the wiring protocol in the following sense. Constructing a wiring of $r$ copies of $P$ together with a local box means that also the parties operating on $P_{\mathrm{loc}}$ need to be assigned to groups. Now since $P_{\mathrm{loc}}$ can be simulated by shared and/or local randomness, any wiring protocol of $P^{\otimes r} \otimes P_{\mathrm{loc}}$ can be translated into an equivalent wiring protocol for $P^{\otimes r}$ which is \textit{stochastic} in the sense that it uses shared and/or local randomness. Conversely, in any wiring protocol which involves shared and/or local randomness, this randomness may be regarded as coming from shared or local coin flips, which can be simulated by an appropriate local box $P_{\mathrm{loc}}$. In conclusion, a deterministic wiring of $P^{\otimes r}\otimes P_{\mathrm{loc}}$ corresponds to a stochastic wiring of $P^{\otimes r}$, and every stochastic wiring arises in this way.

What this implies is that if $P^{\otimes r}$ satisfies LO$^1$, then so does $P^{\otimes r}\otimes P_{\mathrm{loc}}$, and therefore also all deterministic wirings of $P^{\otimes r}\otimes P_{\mathrm{loc}}$, and hence also all stochastic wirings of the original $P^{\otimes r}$. So if $P$ satisfies LO$^\infty$, then the same applies to any box constructed from copies of $P$ via wirings, where these wirings may be dynamic and stochastic. In particular, this even applies to dynamic wirings which make use of shared randomness. This is the most general kind of wiring that we can conceive of.

Surprisingly, in order to show closure under wirings with shared randomness, we have not used convexity of the set $\mathcal{LO}^\infty$. In fact, it remains an open problem whether the set $\mathcal{LO}^\infty$ is convex in all Bell scenarios.

As a simple example application of the results in this section, we show that LO$^\infty$ is violated by \textit{all} extremal boxes in all $(2,2,d)$ and $(2,m,2)$ scenarios. In the $(2,2,d)$ case, this follows from the fact that several copies of any such $d$-outcome extremal box can be wired into an (approximate) PR-box~\cite{BarrettPRA2005}. Then since the PR-box violates LO$^2$, our results show that the original box violates LO$^{2k}$ for some $k$. In fact, if $d$ is even, a single copy of the box is sufficient, and one only needs to apply a coarse-graining to the outcomes of the box, so that it even violates LO$^2$ as well. For the $(2,m,2)$ scenarios, we use the characterization given in~\cite{JM05}(see also~\cite{BP05}), which shows that any extremal box turns into a PR-box upon restricting the measurement choices of each party to only two out of the $m$ possible choices. Since applying this restriction is a trivial kind of wiring, any LO violation by a PR-box translates into an LO violation of any of these extremal boxes.

\section{Constructing UPBs and weak UPBs from LO inequalities}
\label{se:UPB}

Here, exploiting the results of~\cite{UPBBell1,UPBBell2,AAABbook},
we demonstrate how to relate LO inequalities to UPBs, a notion already introduced in the early days of quantum information theory (see
below for the definition). More precisely, every LO inequality within an
$(n,m,2)$ scenario can be turned into an $n$-qubit UPB, while LO inequalities in a general $(n,m,d)$ scenario with $d\geq 3$ can be turned into more general objects called weak UPBs~\cite{AAABbook}.

We begin our detailed considerations by recalling the definition of a UPB
(the definition of a weak UPB will be provided later). To this
end, let us consider a product Hilbert space
\begin{equation}\label{Hspace}
\mathcal{H}=\mathbb{C}^{d_1}\ot\ldots\ot\mathbb{C}^{d_n}
\end{equation}
with $d_i$ denoting the local Hilbert space dimensions at the different sites, and a collection of mutually orthogonal
and normalized product vectors from $\mathcal{H}$,
\[
U=\left\{\ket{\psi_j^{(1)}}\ot\ldots\ot\ket{\psi_j^{(n)}}\right\}_{j=1}^{|U|},
\]
where $\ket{\psi_j^{(i)}}\in\mathbb{C}^{d_i}$. Following~\cite{BennettUPB},
we call $U$ an \textit{unextendible product basis} if (i) it spans a proper
subspace of $\mathcal{H}$, and (ii) $\mathcal{H}$ does not contain any
other product vector orthogonal to all elements of $U$. In other words,
the orthogonal complement $\mathrm{span}(U)^\perp$ in $\mathcal{H}$ is nontrivial and
completely entangled, i.e.~contains no product vectors.

Since UPBs are surprising objects, there has been some interest in characterizing their properties (see e.g.~\cite{Lovasz,UPB2,Bravyi,Duan,Lukasz,Leinaas,NJChen,NJ}). One of their main uses in quantum information theory is that they lead to a general construction of bipartite and multipartite bound entangled states~\cite{BennettUPB}. A state is \textit{bound entangled} if no pure entangled state on any subgroup of the parties can be distilled from it~\cite{HHHBE}. More precisely,
let us denote by $\Pi_U$ the projection onto the subspace spanned by a UPB $U$.
Then, the state
\[
\rho_{U}:=\frac{1}{\dim \mathcal{H}-|U|}(\mathbbm{1}-\Pi_U)
\]
is entangled because its support does not contain any product
vectors. Moreover, it has positive partial transpose with respect to any bipartition. Due to the results of~\cite{HHHBE}, this makes $\rho_U$ bound entangled~\cite{BennettUPB} (see also \cite[XII.I.1]{Hreview}).

We now move on to the construction of UPB's presented
in~\cite{UPBBell1,UPBBell2,AAABbook}. In what follows, we assume
that all vectors are normalized and we also identify every two
vectors differing only by a phase factor, i.e.~any $\ket{\psi}$
and $\alpha\ket{\psi}$ with $|\alpha|=1$ are considered the same
vector. To the Bell scenario $(n,m,d)$ we associate the product
Hilbert space~\eqref{Hspace} with all the local dimensions equal,
i.e., $d_i=d$ for all $i=1,\ldots,n$. This $\mathcal{H}$ is the
smallest Hilbert space that can support projective $d$-outcome
measurements at each site. Then, in each local Hilbert space
$\mathbb{C}^{d}$, we distinguish $m$ different orthogonal bases,
denoted by $B_j=\{\ket{\phi_i^{(j)}}\}_{i=0}^{d-1}$, where
$j=0,\ldots,m-1$. For simplicity, we take these to be the same for
all sites; the generalization to different bases at each site will
be straightforward. We choose them, however, in such way that they
satisfy the following property: \textit{at each site no two
vectors belonging to different bases are orthogonal}. In other
words, if two basis vectors are orthogonal, then they are from the
same basis,
\[
\braket{\phi_i^{(j)}}{\phi_{i'}^{(j')}} = 0 .\quad\Longrightarrow\quad j = j'.
\]
In the following, we
refer to this property as (P). In
the case of qubit Hilbert spaces, i.e.~for $d=2$, this assumption
is always satisfied \cite{UPBBell2}: if a vector from one basis in
$\mathbb{C}^2$ is orthogonal to a vector belonging to the other basis,
then these two bases are the same (up to phase factors).

Consider now an LO inequality in scenario $(n,m,d)$.
To each event $e=(a_1\ldots a_n|x_1\ldots x_n)$
occurring in this inequality, one can associate the product vector
\begin{equation}\label{vector}
\ket{\phi_{a_1}^{(x_1)}}\ot \ldots \ot
\ket{\phi_{a_n}^{(x_n)}}\in\mathcal{H},
\end{equation}
so that for each party $i$, the measurement choice $x_i$ specifies the basis $B_{x_i}$ from which to choose the vector, while the outcome $a_i$ specifies the vector one has to choose from that basis. Applying this procedure to all terms in an LO inequality generates a set of product
vectors of the form~\eqref{vector}. Now pairwise orthogonality of the events in an LO inequality is equivalent to pairwise orthogonality of the associated product vectors: orthogonality of two events means that for some party $i$, the two events have the same measurement choices, but different outcomes. Hence,
at the same party $i$, the corresponding product vectors~\eqref{vector} contain two different
elements of the same local basis, and thus are orthogonal. Consequently, to any LO inequality one can associate a set $S$ of pairwise orthogonal product vectors from $\mathcal{H}$.

Now, if the LO inequality is optimal in the sense that no additional orthogonal events can be added to the left-hand side, then
$S$ is unextendible in the following sense: there is no product
vector $\ket{\varphi}=\ket{\varphi^{(1)}}\ot\ldots\ot\ket{\varphi^{(n)}}\in\mathcal{H}$
with all local components $\ket{\varphi^{(i)}}$ taken from the bases $B_j$
such that $\ket{\varphi}\perp S$. To see this explicitly,
assume that $\ket{\varphi}$ is orthogonal to all elements of $S$. Then,
reversing the above construction, $\ket{\varphi}$ corresponds to an event $(a_1\ldots a_n|x_1\ldots x_n)$ which is orthogonal to all events in the LO inequality, contradicting the fact that the latter is optimal. Following~\cite{AAABbook}, we call a set of orthogonal
vectors $S$ from the $B_j$ which is unextendible in this sense a \textit{weak UPB}. As demonstrated in~\cite{UPBBell2}, the existence of no-signaling violations of the inequality guarantees that this weak UPB is not a complete basis of $\mathcal{H}$.

What we have shown is that every optimal LO inequality gives rise to a weak UPB. As one can see by tracing the steps of this construction backwards, this can also be reversed to turn every weak UPB into an optimal LO inequality~\cite{UPBBell1,UPBBell2,AAABbook}. That the weak UPB is not a complete basis guarantees that the LO inequality is violated by some no-signaling correlations~\cite{UPBBell2}.

Now how do weak UPBs relate to ordinary UPBs? First, any UPB whose local vectors can be grouped into bases having the
property (P) is also a weak UPB with respect to these bases. Second, for $d=2$, every weak UPB is also a UPB. To prove this, assume that $S$ forms a weak UPB
and there is a product vector $\ket{\varphi}=\ket{\varphi^{(1)}}\otimes\ldots\otimes\ket{\varphi^{(n)}}$ orthogonal to $S$. Then if some $\ket{\varphi^{(i)}}$ does not belong to any of the bases $B_j$, it cannot be orthogonal to any local component of any vector in $S$. Therefore, it can likewise be replaced by an arbitrary basis vector from the $B_j$, without destroying the property that $\ket{\varphi}$ is orthogonal to all of $S$. In this way, we can assume without loss of generality that each $\ket{\varphi^{(i)}}$ does belong to some $B_j$. But now since $S$ is assumed to be a weak UPB, it follows that such a $\ket{\varphi}$ cannot exist, and $S$ is a UPB. Third, there are weak UPBs with $d\geq 3$ that are not UPBs; we give an example below.

As a result, any optimal LO inequality within any $(n,m,d)$ scenario
can be turned into a weak UPB in $(\mathbb{C}^d)^{\ot n}$. If the inequality is violated by some no-signaling correlations, then the weak UPB is not just a complete basis. Since LO inequalities can have no-signaling violations only for $n\geq 3$, weak UPBs which are not complete bases can exist only for $n\geq 3$, although there are bipartite UPBs~\cite{BennettUPB,UPB2}. Finally, if $d=2$, then any weak UPB constructed in this way is automatically a UPB.

In particular, all the optimal LO inequalities which we found in the $(4,2,2)$ and $(3,2,3)$ scenarios in Appendix~\ref{Allineqs} give rise to four-qubit
UPBs and three-qutrit weak UPBs, respectively. However, of the four basic types~\ref{permparties} to~\ref{nosignorm} of equivalences between LO inequalities, only the first three have a clear meaning for (weak) UPBs: permuting the sites, permuting the bases at one site, and permuting the elements of one basis at one site. The fourth one does not have an obvious meaning for weak UPBs, and therefore a classification of weak UPBs should only classify LO inequalities with respect to the equivalences~\ref{permparties} to~\ref{permoutcomes}. This is why the results of Appendix~\ref{Allineqs} are not directly applicable for a classification of weak UPBs, but they provide at least, respectively, $35$ and $4$ classes of weak UPBs in the respective $(4,2,2)$ and $(3,2,3)$ scenarios. We now consider these two scenarios as well as the simpler $(3,2,2)$ in a bit more detail. Also, the arXiv version of this paper provides files with a classification of weak UPBs in the $(3,2,3)$ and $(4,2,2)$ scenarios for download.

\paragraph{Scenario $(3,2,2)$.}

The corresponding weak UPBs live on the three-qubit Hilbert space
$\mathcal{H}=(\mathbb{C}^2)^{\ot 3}$. In each local Hilbert space $\mathbb{C}^2$, we distinguish the two bases
\be
\label{B0B1}
B_0=\{\ket{0},\ket{1}\},\qquad B_{1}=\{\ket{e},\ket{e^{\perp}}\},
\ee
with $\ket{e}$ being any vector different from $\ket{0}$ and $\ket{1}$, while
$\ket{e^{\perp}}$ stands for a vector orthogonal to $\ket{e}$. We use notation such as $\ket{0e^\perp}$ as shorthand for $\ket{0}\ot\ket{e^\perp}$.

As discussed in~\cite{UPBBell1}, the only nontrivial LO inequality within this scenario
is the GYNI inequality~\eqref{gyni}. Applying the above prescription for turning an LO inequality into a weak UPB gives the replacements
\begin{align*}
&(000|000) \mapsto\ket{000}, & (110|011) \mapsto\ket{1e^\perp e},\\
&(011|101) \mapsto\ket{e1e^\perp}, & (101|110) \mapsto\ket{e^\perp e1}.
\end{align*}
Therefore GYNI corresponds to the so-called \textit{Shifts} UPB~\cite{BennettUPB},
\[
U_{\mathrm{Shifts}}=\left\{\,\ket{000},\,\ket{1e^{\perp}e},\,\ket{e1e^{\perp}},\,\ket{e^{\perp}e1}\,\right\}.
\]
Indeed, up to permutations of the sites and local unitary operations, this is the unique UPB in $(\mathbb{C}^{2})^{\ot 3}$~\cite{Bravyi}.

\paragraph{Scenario $(4,2,2)$.} Here, the corresponding Hilbert space is $\mathcal{H}=(\mathbb{C}^2)^{\ot 4}$. For the local bases, we again take those given by~\eqref{B0B1}. As already said, all the possible optimal LO inequalities we found within this scenario, in particular
those listed in Tables \ref{ineq422_8} and~\ref{ineq422_>8}, can be turned into four-qubit UPBs. Since $m=2$, each UPB has at most two different bases per site. Some of these were also found in~\cite{UPBBell2}, while most of them are new.
The UPBs obtained in this way have only $8$, $9$, $10$ or $12$ elements, although the minimal size of a four-qubit UPB is $6$~\cite{NJ}. To obtain examples of UPBs of sizes $6$ and $7$, one needs more bases than two for some of the parties~\cite{UPBBell2}. Finally, the $12$-element UPB associated with the largest LO inequality of Table~\ref{ineq422_>8} is of the form $\{\ket{0}\ot\mathcal{B},\ket{1}\ot U\}$ with $\mathcal{B}$
denoting the standard basis in $(\mathbb{C}^{2})^{\ot 3}$, while $U$ is a
three-qubit UPB equivalent to $U_{\mathrm{Shifts}}$.

As a more detailed example, we consider one of the LO
inequalities found within the $(4,2,2)$ scenario, say the fifth
one of Table~\ref{ineq422_8},
\begin{align*}
\label{LOineqEx}
& P(0000|0000)+P(0001|0000)+P(0010|1100)+P(0101|1010)\\[4Pt]
+\; & P(1010|1101)+P(1100|0110)+P(1110|0111)+P(1111|1011)\leq 1.
\end{align*}
With each of the eight events in this inequality, we associate a product vector in $(\mathbb{C}^{2})^{\ot 4}$ as above, which results in the UPB
\[
\left\{\,\ket{0000},\,\ket{0001},\,\ket{ee10},\,\ket{e1e1},\,\ket{e^{\perp}e1e},\,\ket{1e^{\perp}e0},\,\ket{1e^{\perp}e^{\perp}e},\,\ket{e^{\perp}1e^{\perp}e^{\perp}}\,\right\}
.
\]
It is easy to verify by hand that these eight vectors are indeed pairwise orthogonal. Owing to what we have said before, they form a
four-qubit UPB.

\paragraph{Scenario $(3,2,3)$.} The corresponding Hilbert space is $\mathcal{H}=(\C^3)^{\ot 3}$. We choose in $\mathbb{C}^{3}$ two orthonormal bases
\[
B_{0}=\{\ket{0},\ket{1},\ket{2}\},\qquad B_1=\{\ket{e},\ket{e^{\perp}},\ket{e^{\top}}\},
\]
where the three vectors in $B_1$ are such that none of them is orthogonal to any vector in $B_0$.

Then, all the optimal LO inequalities found within this scenario, in particular
those in Table~\ref{ineq323}, can be turned into weak UPBs on $\mathcal{H}$.
However, none of these weak UPBs is actually a UPB.

For example, let us consider the fifth LO inequality in Table~\ref{ineq323},
\begin{align*}
\label{ineq3}
& P(000|000)+P(001|000)+P(002|110)+P(010|000)+P(011|000)+P(012|110)\\[4Pt]
+ \; & P(102|110)+P(112|110)+P(120|011)+P(220|011)+P(221|101)+P(222|101)\leq 1.
\end{align*}
The resulting weak UPB is
\begin{alignat*}{8}
U = \Big\{\, &\ket{000},\,&&\ket{001},\,&&\ket{ee2},\,&&\ket{010},\,&&\ket{011},\,&&\ket{ee^{\perp}2},\,\\[4pt]
&\ket{e^{\perp}e2},\,&&\ket{e^{\perp}e^{\perp}2},\,&&\ket{1e^{\top}e},\,&&\ket{2e^{\top}e},\,&&\ket{e^{\top}2e^{\perp}},\,&&\ket{e^{\top}2e^{\top}}\,\Big\}.
\end{alignat*}
One may verify by hand that this is indeed a weak UPB. That is,
the subspace orthogonal to $U$ does not contain any product vector
$\ket{\alpha\beta\gamma}\in(\mathbb{C}^3)^{\ot 3}$ whose local components
$\ket{\alpha}$, $\ket{\beta}$, and $\ket{\gamma}$ belong to $B_0$ or $B_1$.

Nevertheless, there are product vectors orthogonal to $U$: any product vector $\ket{\alpha\beta\gamma}$ with the properties
\[
\ket{\alpha}\perp\mathrm{span}\{\ket{0},\ket{e}\},\qquad \ket{\beta}\perp\mathrm{span}\{\ket{2},\ket{e^{\top}}\},\qquad \ket{\gamma}\perp\ket{2}
\]
is orthogonal to all elements of $U$. So although $U$ is a weak UPB with respect to the bases $B_0$ and $B_1$, it is not a UPB.

On the other hand, there do exist UPBs on a three-qutrit Hilbert space~\cite{UPB2}, but these do not have property (P). These considerations illustrate the subtle differences between the notions of UPB and weak UPB.

\section{Conclusions and open problems}

The understanding of quantum correlations necessarily requires multipartite information principles. Local orthogonality is a very natural principle with an intrinsic multipartite formulation, which, in spite of its simplicity, reveals a highly non-trivial aspect of quantum correlations and makes it possible to witness the non-physicality of correlations for which all other principles have failed so far.

After introducing the Local Orthogonality principle at the one-copy level, we have shown it to be satisfied by quantum correlations, but violated by some no-signaling correlations. We then explained how this is related to graph theory. This connection with graph theory turns out to be not only conceptually elegant, but also is an indispensable tool for finding violations of the LO principle in practice.

We found that even if a box $P$ satisfies LO, then there may be a number of copies $k$ such that $P^{\otimes k}$ does not satisfy it. For example, this is true for the PR-box with $k=2$. Now since taking such independent copies of a physically realistic box should result in a physically realistic box, we have introduced the LO$^\infty$ principle which postulates LO for any number of copies.

Naively, one might think that LO$^\infty$ could be generalized even further to a principle in which one also postulates LO for any wiring of copies of a box. However, we have shown that this does not give anything better: the set of boxes satisfying LO$^\infty$ is already closed under wirings. Therefore, constructing a wiring cannot lead to new violations of LO$^\infty$. Surprisingly, this closure under wirings holds although the set of boxes satisfying LO$^\infty$ is not known to be convex.

We have also explained how the LO principle relates to unextendible product bases, which gives a useful recipe for generating these.

Because of its intrinsic multipartite formulation, we believe that the LO$^\infty$ principle will play a prominent role in the ongoing quest to understand quantum correlations. It also opens a series of interesting open problems, some of which we explain now.

First, it is known that there are tight Bell inequalities in multipartite scenarios which cannot be violated by quantum correlations~\cite{gyni,UPBBell2,UPBBell2,AAABbook}, and all of these are, in fact, LO inequalities. It is an interesting working conjecture to speculate that any tight Bell inequality without quantum violation is an LO inequality (see \figref{fig:poly}). In particular, due to the equivalence of LO$^1$ and no-signaling in bipartite scenarios, this comprises the hypothesis that any non-trivial tight Bell inequality in a bipartite scenario has quantum violations.

Second, as explained in more detail in~\cite{FLS}, the LO$^\infty$ principle is not strong enough to characterize quantum correlations (this is due to Navascu{\'e}s). In other words, there are still supra-quantum correlations which satisfy LO$^\infty$. There may be other general operations among boxes which go beyond even wirings. Is this the case? If so, can the LO$^\infty$ principle be further generalized by exploiting these? For example, following~\cite{Yan}, it is possible to define a strengthened principle for contextuality scenarios. However, this extension already \textit{assumes} that quantum correlations are physically realizable, which is an undesirable premise. Is there a way around this?

Third, it is a challenging open problem to find out whether the set of boxes satisfying LO$^\infty$ is convex. On a related note, we can ask whether violations of LO$^\infty$ can be \textit{activated}. This means: are there boxes $P_1$ and $P_2$ such that both satisfy LO$^\infty$, but $P_1\otimes P_2$ does not?

\section*{Acknowledgements}

We would like to thank Elie Wolfe for comments.

This research is supported by the Government of Canada
through Industry Canada and by the Province of Ontario through the
Ministry of Economic Development and Innovation, by a grant from
the John Templeton Foundation, by the Excellence Initiative of the
German Federal and State Governments (Grant ZUK 43), by the EU IP
SIQS, by the Spanish CHIST-ERA DIQIP and FIS2010-14830 projects
and a Juan de la Cierva Scholarship, and by the ERC StG PERCENT. ABS was also supported by the AP2009-1174 FPU PhD grant.

\appendix

\section{Classifying LO inequalities}
\label{sec.equiv}

In this appendix, we show how LO inequalities can be
computed and classified in a general Bell scenario. We first discuss how
symmetries can be taken into account to define equivalence classes
of LO inequalities and then list all the relevant LO
inequalities for some scenarios.

The arXiv version of this paper also provides files with the inequalities in the $(3,2,3)$ and $(4,2,2)$ scenarios for download.

\subsection{Defining and computing equivalence classes of LO inequalities}

Starting from the orthogonality graph for a given scenario $(n,m,d)$, as defined in \secref{se:lo&gt}, one can generate a list of all the corresponding LO inequalities by employing the graph-theoretic tools explained in Section~\ref{se:lo&gt}. For sufficiently small scenarios, this computation is feasible with the existing software packages for clique enumeration~\cite{mace,cliquer}, and here we describe the results of our computations along these lines. We write $\mathbf{a}:=a_1\ldots a_n$ and $\mathbf{x}:=x_1\ldots x_n$ as shorthands.

Each LO inequality is of the form
\begin{equation}
\label{LOineqcoeffs}
\sum_{\mathbf{a},\mathbf{x}} c_{\mathbf{a},\mathbf{x}} P(\mathbf{a}|\mathbf{x}) \leq 1 ,
\end{equation}
with $c_{\mathbf{a},\mathbf{x}} \in \{0,1\}$, that is, each LO inequality corresponds simply to a list of the terms which are present, i.e.~the terms for which $c_{\mathbf{a},\mathbf{x}} = 1$. However, for the purpose of understanding the structural aspects of the LO principle, many of these inequalities can be considered equivalent. More concretely, if two inequalities with respective coefficients $c_{\mathbf{a},\mathbf{x}}$ and $c'_{\mathbf{a},\mathbf{x}}$ can be transformed into each other by relabelling the parties or the measurement choices and outcomes, or by making use of the normalization and no-signaling equations, or by combining such transformations, then they really represent different instances of the same basic inequality. So we consider two LO inequalities to be equivalent if one can be transformed into the other under a combination of the following transformations:
\renewcommand{\theenumi}{(\roman{enumi})}
\renewcommand{\labelenumi}{(\roman{enumi})}
\begin{enumerate}
 \item\label{permparties} \textit{Permutation of parties}. For some permutation $\sigma$ of $n$ objects, the $c'_{\mathbf{a},\mathbf{x}}$ of the second inequality can be obtained from the $c_{\mathbf{a},\mathbf{x}}$ of the first inequality as $c'_{\mathbf{a},\mathbf{x}} = c_{\mathbf{a}',\mathbf{x}'}$, where $a'_i = a_{\sigma(i)}$ and $x'_i = x_{\sigma(i)}$.

 \item\label{permsettings} \textit{Relabelling of measurement choices}. For some set of permutations $\sigma_1,\ldots,\sigma_n$ of $m$ objects, the coefficients of the second inequality can be obtained as $c'_{\mathbf{a},\mathbf{x}} = c_{\mathbf{a},\mathbf{x}'}$, where $x'_i = \sigma_i(x_i)$.

 \item\label{permoutcomes} \textit{Relabelling of outcomes}. For some set of permutations $\sigma_{1,1},\ldots,\sigma_{n,m}$ of $d$ objects indexed by parties and measurement choices, the coefficients of the second inequality can be obtained as $c'_{\mathbf{a},\mathbf{x}} = c_{\mathbf{a}',\mathbf{x}}$, where $a'_i = \sigma_{i,x_i}(a_i)$.

 \item\label{nosignorm} \textit{No-signaling and normalization}. The second inequality can be obtained from the first one by adding a linear combination of the no-signaling equations~\eqref{NS-cond} and the normalization condition $\sum_{\mathbf{a}} P(\mathbf{a}|\mathbf{x}) = 1$.
\end{enumerate}

Since the clique enumeration software enumerated \textit{all} cliques in the respective orthogonality graphs, the corresponding sets of LO inequalities had large redundancy in the sense that many inequalities were equivalent to each other under these symmetry transformations. In the following, we describe how we eliminated this redundancy by computing one unique representative of each symmetry class.

First, we considered each inequality for $n$ parties with $r$ terms as an $(r \times 2n)$-matrix. Each row in the matrix corresponds to a term of the inequality by listing the corresponding outcomes and measurement choices $a_1\ldots a_n x_1\ldots x_n$. Two such matrices which differ only by the order of their rows trivially represent the same inequality, and hence we choose the lexicographically smallest ordering as a \textit{normal form} with respect to this equivalence: two matrices represent the same inequality if and only if they have the same normal form. In all subsequent steps, an inequality was always represented as a matrix whose rows are lexicographically ordered. More generally, we always reduced the elimination of equivalences to the computation of normal forms.

Next we eliminated the equivalences under transformations of types~\ref{permparties}--\ref{permoutcomes}, again by computing a normal form with respect to these transformations for each inequality. For each party, the measurement choices were ordered according to their multiplicity, i.e.~according to the number of terms in which they appear. They were then relabelled such that the measurement choice which occurred most often was assigned the lowest label, and so on for the following measurement choices. Similarly, for each party and each measurement choice, the outputs were relabelled according to their multiplicity. Whenever multiple measurement choices or outcomes occurred with the same multiplicity, all possible relabellings were applied, resulting in a list of equivalent inequalities. Next, all possible permutations of the parties were applied, resulting in an even longer list of inequalities. Then again, for each matrix representing an inequality in the list, the rows were ordered lexicographically---corresponding to a permutation of the terms in the inequality---and then the matrices themselves were ordered lexicographically. The first matrix in this reordered list was taken to be the normal form representating the whole equivalence class. The relabellings of measurement choices and outcomes defined in this way are invariant under permutation of parties and terms, since such permutations cannot change the multiplicity of a given measurement choice or outcome. This ensures that the representative is unique. This defines a normal form with respect to the equivalences~\ref{permparties}--\ref{permoutcomes}, as well as an algorithm to compute it. In this way, we eliminated these equivalences using a piece of \textsc{Mathematica} code. This produced a smaller list of inequalities given in the form~\eqref{LOineqcoeffs}.

Finally, we had to eliminate equivalences under transformations which also include those of type~\ref{nosignorm}. To this end, a normal form for general Bell inequalities and a method for computing this normal form had previously been described in~\cite{bancal2010}. This normal form expresses the inequalities in terms of generalized correlators (see Appendix A of~\cite{bancal2010} and also~\cite{tri-box}). A \textsc{Matlab} package for computing this normal form has been developed by Bancal and was kindly provided to us. Although this software is capable of eliminating equivalences of all types~\ref{permparties}--\ref{nosignorm}, our strategy of first eliminating~\ref{permparties}--\ref{permoutcomes} has turned out to be advantageous: in contrast to general Bell inequalities, our LO inequalities are very sparse and all of their coefficients are $0$ or $1$. This is a feature that we have exploited in our \textsc{Mathematica} code, which does not store the inequalities as large arrays of coefficients, but as $(r\times 2n)$-matrices as explained above, which led to a significant speed-up. We applied Bancal's \textsc{Matlab} software to the list of inequalities obtained in the previous step, which resulted in further elimination of equivalences, this time finally under all of~\ref{permparties}--\ref{nosignorm}. In the end, the representative of each equivalence class in its matrix representation was taken to be the first inequality of the class in the sorted output from \textsc{Mathematica}.

\subsection{All LO inequalities for the (3,2,2), (3,2,3) and (4,2,2) scenarios}
\label{Allineqs}

Using the method of the previous subsection, we were able to
completely classify all LO inequalities for scenarios (3,2,3)
and (4,2,2). In the tables below we list the normal form
representative of each of the non-trivial equivalence classes.
Here, an inequality is non-trivial if it can be violated by some
no-signaling box. All the other inequalities are trivial,
i.e.~represent the normalization of probabilities or the
no-signaling condition, and thereby are equivalent
under~\ref{nosignorm} to the trivial inequality $0\leq 0$.

\paragraph{Scenario (3,2,2).}

The GYNI inequality~\eqref{gyni} represents the only class in this
scenario.

\paragraph{Scenario (3,2,3).}

Table~\ref{ineq323} lists the four equivalence classes found for
scenario (3,2,3). These four inequalities correspond,
respectively, to maximal cliques of $12$, $13$, $14$, and $15$
vertices in the orthogonality graph $O_{3,2,3}$.

\definecolor{darkblue}{rgb}{0,0,.4}
\newcommand*\circled[1]{\tikz[baseline=(char.base)]{
            \node[shape=circle,draw,inner sep=2pt,color=darkblue] (char) {#1};}}

\newcounter{ineqnumber}
\begin{table}
\begin{center}
\begin{tabular}{|c|c|c|c|}
\hline&&&\\[-9pt]
\addtocounter{ineqnumber}{1}\raisebox{.5pt}{\circled{\raisebox{-.9pt}{\arabic{ineqnumber}}}}
&
\addtocounter{ineqnumber}{1}\raisebox{.5pt}{\circled{\raisebox{-.9pt}{\arabic{ineqnumber}}}}
&
\addtocounter{ineqnumber}{1}\raisebox{.5pt}{\circled{\raisebox{-.9pt}{\arabic{ineqnumber}}}}
&
\addtocounter{ineqnumber}{1}\raisebox{.5pt}{\circled{\raisebox{-.9pt}{\arabic{ineqnumber}}}}\\[5pt]
\begin{tabular}{c}
$000|000$\\ $001|000$\\ $002|110$\\ $010|000$\\ $011|000$\\
$012|110$\\ $102|110$\\ $112|110$\\ $120|011$\\ $220|011$\\
$221|101$\\ $222|101$
\end{tabular}
&
\begin{tabular}{c}
$000|001$\\ $001|001$\\ $002|111$\\ $010|001$\\ $011|001$\\
$110|010$\\ $120|010$\\ $121|100$\\ $122|100$\\ $210|010$\\
$220|010$\\ $221|100$\\ $222|100$
\end{tabular}
&
\begin{tabular}{c}
$000|000$\\ $001|000$\\ $002|110$\\ $010|000$\\ $011|000$\\
$012|110$\\ $100|000$\\ $101|000$\\ $110|000$\\ $111|000$\\
$120|101$\\ $220|101$\\ $221|011$\\ $222|011$
\end{tabular}
&
\begin{tabular}{c}
$000|000$\\ $001|000$\\ $002|110$\\ $010|000$\\ $011|000$\\
$012|110$\\ $100|000$\\ $101|000$\\ $102|110$\\ $110|000$\\
$111|000$\\ $112|110$\\ $220|011$\\ $221|011$\\ $222|101$
\end{tabular}
\\\hline
\end{tabular}
\end{center}
\caption{The four equivalence classes of LO inequalities in the
$(3,2,3)$ scenario.} \label{ineq323}
\end{table}

\paragraph{Scenario (4,2,2).}

Tables~\ref{ineq422_8} and~\ref{ineq422_>8} list all $35$
equivalence classes found for scenario (4,2,2).
Table~\ref{ineq422_8} contains $30$ inequivalent inequalities with
$8$ terms each. Table~\ref{ineq422_>8} contains $5$ other
inequalities, two of them with $9$ terms, another two with $10$
terms, and a final inequality containing $12$ terms. Taken
together, these are the $35$ classes of non-trivial LO
inequalities in (4,2,2).

\paragraph{Other scenarios.}

These results suggest that the gap between LO$^1$ and no-signaling
increases with the number of parties, in the sense that the number
of classes of optimal LO inequalities grows rapidly. This is why,
for $n>4$ parties, classifying all LO inequalities even in the
simplest scenario $(n,2,2)$ becomes computationally intractable.
Nevertheless, examples of such inequalities for larger $n$ or
larger $m$ and $d$ are known and can be constructed from UPBs (see
\secref{se:UPB}).

\begin{table}[p!]
\floatpagestyle{empty}
\begin{center}
\begin{tabular}{|c|c|c|c|c|c|}
\hline&&&&&\\[-9pt]
\setcounter{ineqnumber}{0}
\addtocounter{ineqnumber}{1}\raisebox{.5pt}{\circled{\raisebox{-.9pt}{\arabic{ineqnumber}}}}
&
\addtocounter{ineqnumber}{1}\raisebox{.5pt}{\circled{\raisebox{-.9pt}{\arabic{ineqnumber}}}}
&
\addtocounter{ineqnumber}{1}\raisebox{.5pt}{\circled{\raisebox{-.9pt}{\arabic{ineqnumber}}}}
&
\addtocounter{ineqnumber}{1}\raisebox{.5pt}{\circled{\raisebox{-.9pt}{\arabic{ineqnumber}}}}
&
\addtocounter{ineqnumber}{1}\raisebox{.5pt}{\circled{\raisebox{-.9pt}{\arabic{ineqnumber}}}}
&
\addtocounter{ineqnumber}{1}\raisebox{.5pt}{\circled{\raisebox{-.9pt}{\arabic{ineqnumber}}}}\\[5pt]
\begin{tabular}{c}
$0000|0000$ \\ $0001|0000$ \\ $0010|0000$ \\ $0100|1011$ \\
$0111|1011$ \\ $1001|0111$ \\ $1010|0111$ \\ $1111|1100$
\end{tabular}
&
\begin{tabular}{c}
$0000|0000$ \\ $0001|0000$ \\ $0010|1100$ \\ $0011|1100$ \\
$1100|0110$ \\ $1101|0110$ \\ $1110|1010$ \\ $1111|1010$
\end{tabular}
&
\begin{tabular}{c}
$0000|0000$ \\ $0001|0000$ \\ $0010|1100$ \\ $0011|1100$ \\
$1100|0110$ \\ $1101|1010$ \\ $1110|0111$ \\ $1111|1011$
\end{tabular}
&
\begin{tabular}{c}
$0000|0000$ \\ $0001|0000$ \\ $0010|1100$ \\ $0011|1100$ \\
$1100|0110$ \\ $1101|1010$ \\ $1110|1010$ \\ $1111|0110$
\end{tabular}
&
\begin{tabular}{c}
$0000|0000$\\ $0001|0000$\\ $0010|1100$\\ $0101|1010$\\
$1010|1101$\\ $1100|0110$\\ $1110|0111$ \\$1111|1011$
\end{tabular}
&
\begin{tabular}{c}
$0000|0000$ \\ $0001|0000$ \\ $0010|1100$ \\ $0101|1010$ \\
$1011|0110$ \\ $1100|0110$ \\ $1110|1010$ \\ $1111|1100$
\end{tabular}
\\\hline&&&&&\\[-9pt]
\addtocounter{ineqnumber}{1}\raisebox{.5pt}{\circled{\raisebox{-.9pt}{\arabic{ineqnumber}}}}
&
\addtocounter{ineqnumber}{1}\raisebox{.5pt}{\circled{\raisebox{-.9pt}{\arabic{ineqnumber}}}}
&
\addtocounter{ineqnumber}{1}\raisebox{.5pt}{\circled{\raisebox{-.9pt}{\arabic{ineqnumber}}}}
&
\addtocounter{ineqnumber}{1}\raisebox{.5pt}{\circled{\raisebox{-.9pt}{\arabic{ineqnumber}}}}
&
\addtocounter{ineqnumber}{1}\raisebox{.5pt}{\circled{\raisebox{-.9pt}{\arabic{ineqnumber}}}}
&
\addtocounter{ineqnumber}{1}\raisebox{.5pt}{\circled{\raisebox{-.9pt}{\arabic{ineqnumber}}}}\\[5pt]
\begin{tabular}{c}
$0000|0000$ \\ $0001|0000$ \\ $0010|1100$ \\ $0101|1010$ \\
$1011|0110$ \\ $1100|1010$ \\ $1110|0110$ \\ $1111|1100$
\end{tabular}
&
\begin{tabular}{c}
$0000|0000$\\ $0001|0010$\\ $0010|1100$\\ $0011|1110$\\
$1100|0101$\\ $1101|0111$\\ $1110|1001$\\ $1111|1011$
\end{tabular}
&
\begin{tabular}{c}
$0000|0000$\\ $0001|0010$\\ $0010|1100$\\ $0011|1110$\\
$1100|0101$\\ $1101|1011$\\ $1110|1001$\\ $1111|0111$
\end{tabular}
&
\begin{tabular}{c}
$0000|0000$\\ $0001|0010$\\ $0010|1100$\\ $0100|1001$\\
$1001|0101$\\ $1100|1011$\\ $1101|0111$\\ $1111|1110$
\end{tabular}
&
\begin{tabular}{c}
$0000|0000$\\ $0001|0010$\\ $0010|1100$\\ $0100|1001$\\
$1011|0100$\\ $1101|0111$\\ $1110|1010$\\ $1111|1110$
\end{tabular}
&
\begin{tabular}{c}
$0000|0000$ \\ $0001|0010$ \\ $0010|1100$ \\ $0100|1001$ \\
$1011|1110$ \\ $1101|1011$ \\ $1110|0101$ \\ $1111|0111$
\end{tabular}
\\\hline&&&&&\\[-9pt]
\addtocounter{ineqnumber}{1}\raisebox{.5pt}{\circled{\raisebox{-.9pt}{\arabic{ineqnumber}}}}
&
\addtocounter{ineqnumber}{1}\raisebox{.5pt}{\circled{\raisebox{-.9pt}{\arabic{ineqnumber}}}}
&
\addtocounter{ineqnumber}{1}\raisebox{.5pt}{\circled{\raisebox{-.9pt}{\arabic{ineqnumber}}}}
&
\addtocounter{ineqnumber}{1}\raisebox{.5pt}{\circled{\raisebox{-.9pt}{\arabic{ineqnumber}}}}
&
\addtocounter{ineqnumber}{1}\raisebox{.5pt}{\circled{\raisebox{-.9pt}{\arabic{ineqnumber}}}}
&
\addtocounter{ineqnumber}{1}\raisebox{.5pt}{\circled{\raisebox{-.9pt}{\arabic{ineqnumber}}}}\\[5pt]
\begin{tabular}{c}
$0000|0000$ \\ $0001|0010$ \\ $0010|1100$ \\ $0101|1000$ \\
$1011|0100$ \\ $1100|0110$ \\ $1110|1010$ \\ $1111|1110$
\end{tabular}
&
\begin{tabular}{c}
$0000|0000$ \\ $0001|0010$ \\ $0010|1100$ \\ $0101|1000$ \\
$1011|0100$ \\ $1100|1010$ \\ $1110|0110$ \\ $1111|1110$
\end{tabular}
&
\begin{tabular}{c}
$0000|0000$ \\ $0001|0010$ \\ $0010|1100$ \\ $0101|1000$ \\
$1011|1110$ \\ $1100|1010$ \\ $1110|0110$ \\ $1111|0100$
\end{tabular}
&
\begin{tabular}{c}
$0000|0000$ \\ $0001|0010$ \\ $0100|1010$ \\ $0101|1000$ \\
$1010|0110$ \\ $1011|0100$ \\ $1110|1100$ \\ $1111|1110$
\end{tabular}
&
\begin{tabular}{c}
$0000|0000$ \\ $0001|0010$ \\ $0100|1010$ \\ $0101|1000$ \\
$1010|1100$ \\ $1011|0100$ \\ $1110|0110$ \\ $1111|1110$
\end{tabular}
&
\begin{tabular}{c}
$0000|0000$ \\ $0001|0010$ \\ $0100|1010$ \\ $0111|1000$ \\
$1001|0100$ \\ $1010|0110$ \\ $1110|1100$ \\ $1111|1110$
\end{tabular}
\\\hline&&&&&\\[-9pt]
\addtocounter{ineqnumber}{1}\raisebox{.5pt}{\circled{\raisebox{-.9pt}{\arabic{ineqnumber}}}}
&
\addtocounter{ineqnumber}{1}\raisebox{.5pt}{\circled{\raisebox{-.9pt}{\arabic{ineqnumber}}}}
&
\addtocounter{ineqnumber}{1}\raisebox{.5pt}{\circled{\raisebox{-.9pt}{\arabic{ineqnumber}}}}
&
\addtocounter{ineqnumber}{1}\raisebox{.5pt}{\circled{\raisebox{-.9pt}{\arabic{ineqnumber}}}}
&
\addtocounter{ineqnumber}{1}\raisebox{.5pt}{\circled{\raisebox{-.9pt}{\arabic{ineqnumber}}}}
&
\addtocounter{ineqnumber}{1}\raisebox{.5pt}{\circled{\raisebox{-.9pt}{\arabic{ineqnumber}}}}\\[5pt]
\begin{tabular}{c}
$0000|0000$ \\ $0001|0110$ \\ $0010|1100$ \\ $0011|1010$ \\
$1100|0001$ \\ $1101|0111$ \\ $1110|1101$ \\ $1111|1011$
\end{tabular}
&
\begin{tabular}{c}
$0000|0000$ \\ $0001|0110$ \\ $0010|1100$ \\ $0011|1010$ \\
$1100|0110$ \\ $1101|0000$ \\ $1110|1010$ \\ $1111|1100$
\end{tabular}
&
\begin{tabular}{c}
$0000|0000$\\ $0001|0110$\\ $0010|1100$\\ $0011|1010$\\
$1100|1010$\\ $1101|0000$\\ $1110|0110$\\ $1111|1100$
\end{tabular}
&
\begin{tabular}{c}
$0000|0000$ \\ $0001|0110$ \\ $0010|1100$ \\ $0011|1010$ \\
$1100|1010$ \\ $1101|1100$ \\ $1110|0110$ \\ $1111|0000$
\end{tabular}
&
\begin{tabular}{c}
$0000|0000$ \\ $0001|0110$ \\ $0010|1100$ \\ $0110|0101$ \\
$1011|0010$ \\ $1100|0001$ \\ $1101|0111$ \\ $1111|1011$
\end{tabular}
&
\begin{tabular}{c}
$0000|0000$ \\ $0001|0110$ \\ $0010|1100$ \\ $0111|1010$ \\
$1001|0000$ \\ $1100|0110$ \\ $1110|1010$ \\ $1111|1100$
\end{tabular}
\\\hline&&&&&\\[-9pt]
\addtocounter{ineqnumber}{1}\raisebox{.5pt}{\circled{\raisebox{-.9pt}{\arabic{ineqnumber}}}}
&
\addtocounter{ineqnumber}{1}\raisebox{.5pt}{\circled{\raisebox{-.9pt}{\arabic{ineqnumber}}}}
&
\addtocounter{ineqnumber}{1}\raisebox{.5pt}{\circled{\raisebox{-.9pt}{\arabic{ineqnumber}}}}
&
\addtocounter{ineqnumber}{1}\raisebox{.5pt}{\circled{\raisebox{-.9pt}{\arabic{ineqnumber}}}}
&
\addtocounter{ineqnumber}{1}\raisebox{.5pt}{\circled{\raisebox{-.9pt}{\arabic{ineqnumber}}}}
&
\addtocounter{ineqnumber}{1}\raisebox{.5pt}{\circled{\raisebox{-.9pt}{\arabic{ineqnumber}}}}\\[5pt]
\begin{tabular}{c}
$0000|0000$ \\ $0001|0110$ \\ $0010|1100$ \\ $0111|1010$ \\
$1001|0000$ \\ $1100|1010$ \\ $1110|0110$ \\ $1111|1100$
\end{tabular}
&
\begin{tabular}{c}
$0000|0000$ \\ $0001|0110$ \\ $0010|1100$ \\ $0111|1100$ \\
$1001|0000$ \\ $1100|1010$ \\ $1110|0110$ \\ $1111|1010$
\end{tabular}
&
\begin{tabular}{c}
$0000|0000$ \\ $0001|0110$ \\ $0110|0011$ \\ $0111|0101$ \\
$1000|0110$ \\ $1001|0000$ \\ $1110|0101$ \\ $1111|0011$
\end{tabular}
&
\begin{tabular}{c}
$0000|0000$ \\ $0001|0110$ \\ $0110|0011$ \\ $0111|1011$ \\
$1000|0110$ \\ $1001|0000$ \\ $1110|0101$ \\ $1111|1101$
\end{tabular}
&
\begin{tabular}{c}
$0000|0000$ \\ $0001|0110$ \\ $0110|1011$ \\ $0111|1101$ \\
$1000|0110$ \\ $1001|0000$ \\ $1110|1101$ \\ $1111|1011$
\end{tabular}
&
\begin{tabular}{c}
$0000|0000$ \\ $0011|0100$ \\ $0101|1000$ \\ $0110|1100$ \\
$1001|0010$ \\ $1010|0110$ \\ $1100|1010$ \\ $1111|1110$
\end{tabular}
\\\hline
\end{tabular}
\end{center}
\caption{Equivalence classes $8$-term LO inequalities in the
$(4,2,2)$ scenario.} \label{ineq422_8}
\end{table}

\begin{table}
\begin{center}
\begin{tabular}{|c|c|c|c|c|}
\hline&&&&\\[-9pt]
\addtocounter{ineqnumber}{1}\raisebox{.5pt}{\circled{\raisebox{-.9pt}{\arabic{ineqnumber}}}}
&
\addtocounter{ineqnumber}{1}\raisebox{.5pt}{\circled{\raisebox{-.9pt}{\arabic{ineqnumber}}}}
&
\addtocounter{ineqnumber}{1}\raisebox{.5pt}{\circled{\raisebox{-.9pt}{\arabic{ineqnumber}}}}
&
\addtocounter{ineqnumber}{1}\raisebox{.5pt}{\circled{\raisebox{-.9pt}{\arabic{ineqnumber}}}}
&
\addtocounter{ineqnumber}{1}\raisebox{.5pt}{\circled{\raisebox{-.9pt}{\arabic{ineqnumber}}}}\\[5pt]
\begin{tabular}{c}
$0000|0000$\\ $0001|0000$\\ $0010|0000$\\ $0011|1100$\\
$0100|0001$\\ $1000|0110$\\ $1001|0000$\\ $1110|0101$\\
$1111|1011$
\end{tabular}
&
\begin{tabular}{c}
$0000|0001$\\ $0010|0100$\\ $0011|1000$\\ $0100|1000$\\
$0101|0010$\\ $1000|0010$\\ $1001|0100$\\ $1110|0001$\\
$1111|1111$
\end{tabular}
&
\begin{tabular}{c}
$0000|0000$\\ $0001|0000$\\ $0010|0000$\\ $0011|0000$\\
$0100|0000$\\ $0101|1010$\\ $1000|0100$\\ $1001|0010$\\
$1110|1001$\\ $1111|0111$
\end{tabular}
&
\begin{tabular}{c}
$0000|0000$\\ $0001|0000$\\ $0010|0000$\\ $0011|1100$\\
$0100|0001$\\ $0110|1000$\\ $1000|0101$\\ $1010|0000$\\
$1101|0110$\\ $1111|1011$
\end{tabular}
&
\begin{tabular}{c}
$0000|0000$\\ $0001|0000$\\ $0010|0000$\\ $0011|0000$\\
$0100|0000$\\ $0101|0000$\\ $0110|0000$\\ $0111|0000$\\
$1000|0000$\\ $1001|0110$\\ $1110|0011$\\ $1111|0101$
\end{tabular}
\\\hline
\end{tabular}
\end{center}
\caption{Equivalence classes of LO inequalities with more than $8$
terms in the $(4,2,2)$ scenario.} \label{ineq422_>8}
\end{table}

\section{Shannon capacity of the PR-box graph and box packing}
\label{boxpacking}

In this section, we discuss how to think of the Shannon capacity of the non-orthogonality graph of the PR-box drawn in \figref{circulant}; in the following, we simply write $G$ for this graph. We relate its Shannon capacity to the problem of finding optimal packings of boxes in a $k$-dimensional torus.

The Shannon capacity $\Theta(G)$ is the limit of $\sqrt[k]{\alpha_k}$ as $k\to\infty$, where $\alpha_k=\alpha(G^{\boxtimes k})$ is the independence number of the strong product of $k$ copies of $G$.
Let us write $V(G) := \left\{ 0, 1,\ldots , 7\right\}$ for the vertices of $G$. Then two tuples
\[
v = (v_1, \ldots, v_k),\qquad w=(w_1, \ldots, w_k)
\]
in $V^k$ are non-adjacent as vertices in $G^{\boxtimes k}$ if and only if $|v_i - w_i| \geq 3$ for all $i$, where the difference $v_i-w_i$ is taken modulo 8. This condition can be reformulated as $||v-w||_\infty \geq 3$, where we define $||v-w||_\infty := \max_{i} |v_i - w_i|$.

This can be interpreted in the following way: a vertex $v \in G^{\boxtimes k}$ can be represented as an $n$-dimensional cube with sides of length $3$ centered around the point $v \in (\mathbb{Z}/{8\mathbb{Z}})^k \subset [0,8]^k$. Here, we think of $[0,8]^k$ as a hypercube with opposite sides identified, so that it becomes an $k$-dimensional torus. It has $8^k$ points with integer coordinates. These integer points form the lattice $(\mathbb{Z}/8\mathbb{Z})^k\subset [0,8]^k$. We require each cube to be centered around one of these integer points. In the following, we speak of such cubes as ``boxes''.

\begin{figure}
\centering \subfigure[]{ \label{packing1D}
\includegraphics[scale=0.6]{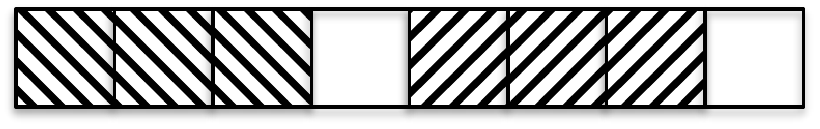}
}\hspace{1.5cm} \subfigure[]{ \label{packing2D}
\includegraphics[scale=0.6]{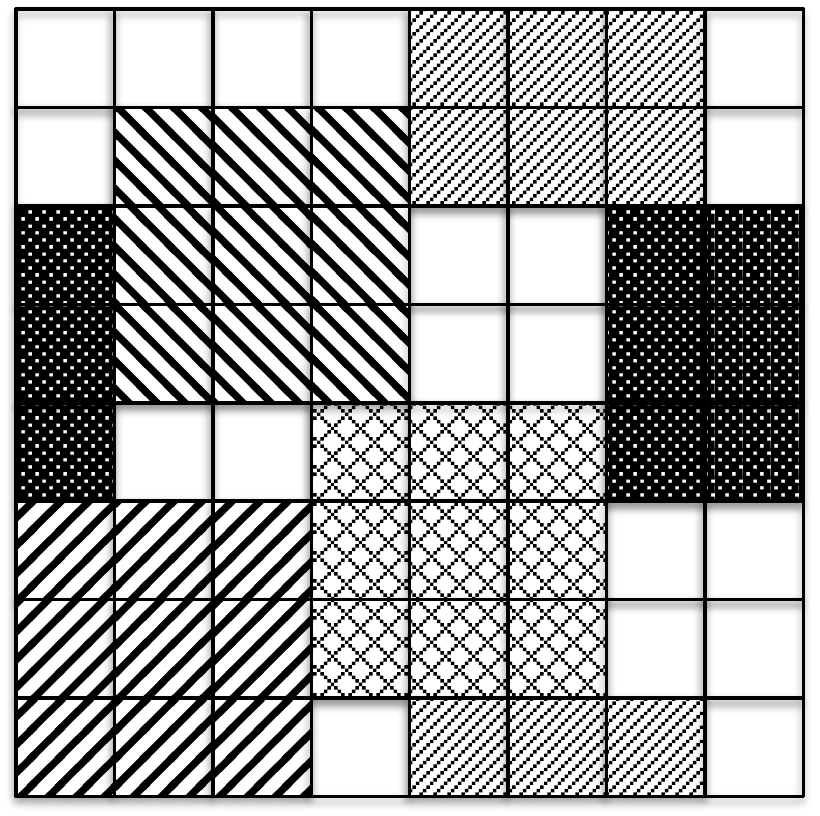}
} \caption{\subref{packing1D} Optimal one-dimensional packing proving that
$\alpha\!\left(G^{\boxtimes 1}\right)=2$ and that $\Theta(G) \geq
2$. \subref{packing2D} Optimal two-dimensional packing proving that
$\alpha\!\left(G^{\boxtimes 2} \right)=5$ and that $\Theta(G) \geq
\sqrt{5}$.}
\label{packingfig} 
\end{figure}

Then two vertices are non-adjacent if an only if the corresponding boxes do not overlap. In particular, finding an independent set in $G^{\boxtimes k}$ becomes equivalent to finding a set of non-overlapping boxes in our torus. See \figref{packingfig} for an illustration. We translated all boxes by $\tfrac{1}{2}$ in each direction to simplify the visualization: each box now occupies three elementary squares in each direction.

It is in fact not necessary to require that the coordinates of the vertices of the hypercubes have integer values: given any non-overlapping configuration characterized by a set of hypercubes $H_i$ described by their ``lowest left'' vertex $x_i \in [0,8]^k$, the configuration of hypercubes $\tilde{H}_i$ with ``lowest left'' vertex $\tilde{x}_i = \lfloor{x_i}\rfloor$ is also non-overlapping. Computing the independence number $\alpha_k$ therefore boils down to asking the following:
\begin{quote}
\textbf{Box packing problem:} How many $k$-dimensional boxes of side length $\tfrac{3}{8}$ fit into the $k$-dimensional torus $[0,1]^k$ (with opposite faces identified)?
\end{quote}
Unfortunately, while it is easy to find optimal box packings in dimensions $k=1$ and $k=2$ as illustrated in \figref{packingfig}, the problem quickly becomes computationally intractable as $k$ increases. Our results of Section~\ref{S_capacities} show that LO$^\infty$ recovers Tsirelson's bound if the number of boxes scales like $(4(2-\sqrt{2}))^k$.

\small
\bibliographystyle{unsrt}
\bibliography{locorth5}

\end{document}